\documentclass[AMA,STIX1COL]{WileyNJD-v2}
\usepackage{moreverb}
\usepackage[caption=false]{subfig}

\def\artanh{\operatorname{artanh}}

\newcommand\BibTeX{{\rmfamily B\kern-.05em \textsc{i\kern-.025em b}\kern-.08em
T\kern-.1667em\lower.7ex\hbox{E}\kern-.125emX}}

\articletype{RESEARCH ARTICL}%

\received{<day> <Month>, <year>}
\revised{<day> <Month>, <year>}
\accepted{<day> <Month>, <year>}

\begin{document}

\title{Eigenvalue Mapping-based Discretization of the Generalized Super-Twisting Algorithm \protect\thanks{This is an example for title footnote.}}


\author[4]{Ningning Ding}
\author[1,2,3]{Yuangui Tang}

\authormark{DING \textsc{et al}}

\address[1]{\orgdiv{Shenyang Institute of Automation}, \orgname{Chinese Academy of Sciences}, \orgaddress{\state{Shenyang}, \country{P.R.China}}}

\address[2]{\orgdiv{Institutes for Robotics and Intelligent Manufacturing}, \orgname{Chinese Academy of Sciences}, \orgaddress{\state{Shenyang}, \country{P.R.China}}}

\address[3]{\orgname{Key Laboratory of Marine Robotics}, \orgaddress{\state{Shenyang}, \country{P.R.China}}}

\address[4]{\orgname{University of Chinese Academy of Sciences}, \orgaddress{\state{Beijing}, \country{P.R.China}}}

 \corres{Yuangui Tang, Shenyang Institute of Automation, Chinese Academy of Sciences, P.R.China.
\email{tyg@sia.cn}}


\abstract[Abstract]
{
In this paper, an eigenvalue mapping-based discretization method is applied to discretize the generalized super-twisting algorithm. The existing eigenvalue mapping is extended to the complex domain which greatly enlarges the range of parameter selection. Furthermore, we present the clue to find new eigenvalue mapping functions (EMFs). One new hybrid EMF and three brand-new EMFs are proposed in this paper. In contrast to the conventional methods, the proposed discretization method totally avoids the discretization chattering and the control precision is enhanced in terms of the steady-state error. Besides, the control precision is insensitive to the overestimation of the control gains, which benefits the gain tuning of the controller in practice. Simulation examples verify the effectiveness and superiority of the proposed discretization methodology.
}


\jnlcitation{\cname{%
\author{Williams K.}, 
\author{B. Hoskins}, 
\author{R. Lee}, 
\author{G. Masato}, and 
\author{T. Woollings}} (\cyear{2016}), 
\ctitle{A regime analysis of Atlantic winter jet variability applied to evaluate HadGEM3-GC2}, \cjournal{Q.J.R. Meteorol. Soc.}, \cvol{2017;00:1--6}.}

\maketitle

\footnotetext{\textbf{Abbreviations:} ANA, anti-nuclear antibodies; APC, antigen-presenting cells; IRF, interferon regulatory factor}

\section{Introduction}\label{sec1}
The super twisting algorithm (STA) has been widely used in the field of robust nonlinear control and observation in recent years.\cite{levant2003higher,cao2022adaptive,zhang2022robust} It is effective in suppressing the chattering while at the same time preserving the good properties of the first-order sliding mode control. As the second-order sliding mode control (SMC), it only requires the sliding variable's information, making it very popular. To improve the robustness and convergence velocity of STA when the system state is far away from the origin, Moreno et al. \cite{moreno2009linear} proposed the generalized super-twisting algorithm (GSTA), which includes additional linear correction terms on the basis of the standard STA. Besides, GSTA can also handle state-dependent disturbances, not just time-dependent disturbances.\cite{castillo2018super,borlaug2022generalized}

Since the common design process of SMC takes place in the continuous time domain and is then implemented in a digital environment, some kind of discretization is necessary. The most commonly applied discretization approach is the explicit Euler method.\cite{yan2015discretization,brogliato2021digital} However, applying this approach to discontinue algorithms such as GSTA yields discretization chattering, i.e., self-sustaining oscillations in the state variables and input, which diminishes the control performance.\cite{yu2003discretization} In recent years, several chattering-free discretization methods have been successfully implemented in STA and related algorithms. An implicit Euler discretization of the STA is proposed in Reference \citenum{brogliato2019implicit}, which suppressed the numerical chattering but achieved only the standard first-order accuracy of SMC. Besides, it cannot provide stability for the same class of disturbances as in the continuous-time case. Furthermore, various other discrete-time variants of the STA that are based on a pseudo-linear system representation are presented in Reference \citenum{koch2019discrete, koch2019discretization}, which is insensitive to an overestimation of the control gains. This eigenvalue-based discretization method is also extended to adaptive gain super-twisting algorithms.\cite{eisenzopf2021adaptive} A semi-implicit Euler method is proposed for STA in Reference \citenum{xiong2021discrete}, which can provide higher asymptotic accuracy and is totally independent of the gain-overestimation. In Reference \citenum{andritsch2023modified}, a modified implicit discretization of STA is proposed. Compared with previously published methods, the proposed method best resembles several properties of the continuous-time controller. Furthermore, the steady-state error is shown to be independent of the controller gains. A new discrete-time version of the STA motivated by the variational framework is presented in Reference \citenum{weissenberger2022discretization}. The proposed variational approach exhibits a great tracking of the continuous-time energy decay rate and gives a good approximation of the continuous-time system. A minimum operator-based approach is proposed for a discrete-time super-twisting-like algorithm.\cite{prasun2023minimum}
It has been proven that this method achieves finite-time stability in the unperturbed case and finite-time input-to-state stability in the perturbed case.

To the best knowledge of the authors, the discretization of GSTA has not yet been addressed in the literature. In this work, the eigenvalue mapping-based discretization method \cite{koch2019discrete} is extended to GSTA considering its simplicity and promising performance. We first established the pseudo-linear representation of the continuous-time closed-loop dynamics derived by GSTA controller and plant dynamics. Then, the eigenvalue mapping from the continuous-time domain to the discrete-time domain is established. We explore the property of existing eigenvalue mapping functions (EMFs) and give the clue to find new EMFs. One new hybrid EMF and three brand-new EMFs are proposed in this paper. Finally, it is demonstrated that the proposed EMFs achieve better control performance than the original EMFs and other discretization methods.

The paper is structured as follows: Section 2 introduces preliminaries. In Section 3, the eigenvalue mapping-based discretization method is extended to GSTA. Section 4 illustrates the effectiveness of the proposed scheme through simulations and comparisons. Finally, Section 5 concludes the paper.


\section{PRELIMINARIES}
Consider the following perturbed first-order system
\begin{eqnarray}
    \dot{x}&=&u+\varphi, \nonumber\\
    \dot{\varphi}&=&\Delta (t).
\label{eq_con_dy_sys}
\end{eqnarray}

\noindent where \(x\in \mathbb{R}\) is the state variable, \(u\in \mathbb{R}\) is the control input and \(\varphi \in \mathbb{R}\) represents the unknown external disturbance with a known Lipschitz constant \emph{L}, i.e., \({{\sup }_{t\ge 0}}\left| \Delta (t) \right|<L\).

The generalized super-twisting controller (GSTC) is given by
\begin{eqnarray}
    u&=&-{{k}_{1}}\left[{{\mu }_{1}}{{\left| x \right|}^{\frac{1}{2}}}sign(x)+{{\mu }_{2}}x\right]+\nu, \nonumber\\
   \dot{\nu}&=&-{{k}_{2}}\left[\tfrac{1}{2}\mu _{1}^{2}sign(x)+\tfrac{3}{2}{{\mu }_{1}}{{\mu }_{2}}{{\left| x \right|}^{\frac{1}{2}}}sign(x)+\mu _{2}^{2}x\right],
\label{eq_con_controller}
\end{eqnarray}

\noindent where \(\nu \in \mathbb{R}\) is the controller state variable, \({{k}_{1}}\), \({{k}_{2}}\), \({{\mu }_{1}}\) and \({{\mu }_{2}}\) are the controller gains to be chosen. Note that for the choice \({{\mu }_{1}}=1\), \({{\mu }_{2}}=0\), the structure of the standard STA is recovered. Additionally, the controller (\ref{eq_con_controller}) reduces to the classical PI controller when ${{\mu }_{1}}=0$ and ${{\mu }_{2}}>0$.

Substituting the control scheme (\ref{eq_con_controller}) into system (\ref{eq_con_dy_sys}) leads to the closed-loop dynamics
\begin{eqnarray}
   \dot{x}&=&-{{k}_{1}}\left[{{\mu }_{1}}{{\left| x \right|}^{\frac{1}{2}}}sign(x)+{{\mu }_{2}}x\right]+\nu +\varphi(t), \nonumber\\
   \dot{\nu }&=&-{{k}_{2}}\left[\tfrac{1}{2}\mu _{1}^{2}sign(x)+\tfrac{3}{2}{{\mu }_{1}}{{\mu }_{2}}{{\left| x \right|}^{\frac{1}{2}}}sign(x)+\mu _{2}^{2}x\right].
\label{closed_loop_con}
\end{eqnarray}

It has been shown in References \citenum{moreno2009linear} and \citenum{castillo2018super}, that a proper choice of control gains ensures finite time stability of the origin despite the disturbance $\varphi(t)$. Solutions of the system (\ref{closed_loop_con}) containing the discontinuous sign function are understood in the sense
of Filippov.\cite{filippov2013differential}

As nowadays controllers are typically implemented in a digital environment, it's necessary to consider the discretization of the continuous controller (\ref{eq_con_controller}) and plant dynamics (\ref{eq_con_dy_sys}). The controller (\ref{eq_con_controller}) discretized by the commonly used explicit Euler method with sampling time $h$ is
\begin{eqnarray}
   {{u}_{k}}&=&-{{k}_{1}}\left[{{\mu }_{1}}{{\left| {{x}_{k}} \right|}^{\frac{1}{2}}}sign({{x}_{k}})+{{\mu }_{2}}{{x}_{k}}\right]+{{\nu }_{k}}, \nonumber\\
   {{\nu}_{k+1}}&=&{{\nu}_{k}}-h{{k}_{2}}\left[\tfrac{1}{2}\mu _{1}^{2}sign({{x}_{k}})+\tfrac{3}{2}{{\mu }_{1}}{{\mu }_{2}}{{\left| {{x}_{k}} \right|}^{\frac{1}{2}}}sign({{x}_{k}})+\mu _{2}^{2}{{x}_{k}}\right],
\label{discrete_euler}
\end{eqnarray}

\noindent where \({{u}_{k}}:=u(kh)\), \({{\nu}_{k}}:=\nu(kh)\) and \({{x}_{k}}:=x(kh)\) with $k=0,1,2,...$

After implementing the controller via a zero-order hold element, a discretization of the system (\ref{eq_con_dy_sys}) with constant sampling time $h$ is given by
\begin{eqnarray}
   {{x}_{k+1}}&=&{{x}_{k}}+h({{u}_{k}}+{{\varphi }_{k}})+\frac{{{h}^{2}}}{2}{{\Delta }_{1,k}}, \nonumber\\
    {{\varphi }_{k+1}}&=&{{\varphi }_{k}}+h{{\Delta }_{2,k}},
\label{zoh_dynamics}
\end{eqnarray}
\noindent where ${{\varphi }_{k}}:=\varphi (kh)$, \({{\Delta }_{1,k}},{{\Delta }_{2,k}}\in [-L,L]\) represents the effect of the external disturbance. Noted that for a continued disturbance \(\Delta (t)\), \({{\Delta }_{1,k}}=\Delta ({{\xi }_{1,k}})\) and \({{\Delta }_{2,k}}=\Delta ({{\xi }_{2,k}})\) for some ${{\xi }_{1,k}},{{\xi }_{2,k}}\in [kh,(k+1)h]$.

The closed-loop system resulting from explicit Euler discretized controller (\ref{discrete_euler}) and the discrete plant dynamics (\ref{zoh_dynamics}) is governed by the recursions
\begin{eqnarray}
     {{x}_{1,k+1}}&=&{{x}_{1,k}}-h{{k}_{1}}\left( {{\mu }_{1}}{{\left| {{x}_{1,k}} \right|}^{\frac{1}{2}}}sign({{x}_{1,k}})+{{\mu }_{2}}{{x}_{1,k}} \right)+h{{x}_{2,k}}+\frac{{{h}^{2}}}{2}{{\Delta }_{1,k}}, \nonumber\\
    {{x}_{2,k+1}}&=&{{x}_{2,k}}-h{{k}_{2}}\left[ \tfrac{1}{2}\mu _{1}^{2}sign({{x}_{1,k}})+\tfrac{3}{2}{{\mu }_{1}}{{\mu }_{2}}{{\left| {{x}_{1,k}} \right|}^{\frac{1}{2}}}sign({{x}_{1,k}})+\mu _{2}^{2}{{x}_{1,k}} \right]+h{{\Delta }_{2,k}},
\label{closed_loop_explicit_euler}
\end{eqnarray}

\noindent with \({{x}_{1,k}}:={{x}_{k}}\) and \({{x}_{2,k}}:={{\nu}_{k}}+{{\varphi }_{k}}\).

For the STA, it has been proven that even in the unperturbed case, the state variables \({{x}_{1,k}}\) and \({{x}_{2,k}}\) of the closed-loop system (\ref{closed_loop_explicit_euler}) do not converge to the origin and eventually exhibit a periodic motion, i.e., the discretization chattering.\cite{yan2020euler} Simulation results show that this is also true for the explicit Euler discretized GSTA.

For the explicit Euler method, it's very easy to implement but there exists severe discretization chattering. Thus proper discretization method should be considered. To this end, an eigenvalue mapping-based discretization method is implemented on the GSTA, which entirely avoids discretization chattering and provides higher precision.

\section{PROPOSED DISCRETIZATION SCHEME}

In order to derive discrete-time versions of the GSTA, an alternative representation of the sign function is used here.\cite{reichhartinger2018arbitrary}
\begin{equation}
sign(x)=
\begin{cases}
x{{\left| x \right|}^{-1}}& \text{ $x\ne 0$ } \\
0& \text{ $ x=0 $ }
\end{cases}
\label{eq_sign_alter}
\end{equation}

Substituting (\ref{eq_sign_alter}) into (\ref{closed_loop_con}), the closed-loop system (\ref{closed_loop_con}) can be represented in a so-called pseudo linear form
\begin{equation}
\dot{x}=\boldsymbol{M}({{x}_{1}})x+\boldsymbol{c}\Delta (t),
\label{eq_continue_matrix}
\end{equation}

\noindent where \(\boldsymbol{x}={{[{{x}_{1}},\ {{x}_{2}}]}^{T}}\), \(\boldsymbol{c}={{[0,\ 1]}^{T}}\) and
\begin{equation}
\boldsymbol{M}({{x}_{1}})=\left[ \begin{matrix}
   -{{k}_{1}}{{\mu }_{1}}{{\left| {{x}_{1}} \right|}^{-\frac{1}{2}}}-{{k}_{1}}{{\mu }_{2}} & 1  \\
   -\frac{1}{2}{{k}_{2}}\mu _{1}^{2}{{\left| {{x}_{1}} \right|}^{-1}}-\frac{3}{2}{{k}_{2}}{{\mu }_{1}}{{\mu }_{2}}{{\left| {{x}_{1}} \right|}^{-\frac{1}{2}}}-{{k}_{2}}\mu _{2}^{2} & 0  \\
\end{matrix} \right].
\end{equation}

The characteristic polynomial of \(\boldsymbol{M}({{x}_{1}})\) is given by
\begin{equation}
\\w(\lambda )={{\lambda }^{2}}+\left({{k}_{1}}{{\mu }_{1}}{{\left| {{x}_{1}} \right|}^{-\frac{1}{2}}}+{{k}_{1}}{{\mu }_{2}}\right)\lambda +\tfrac{1}{2}{{k}_{2}}\mu _{1}^{2}{{\left| {{x}_{1}} \right|}^{-1}}+\tfrac{3}{2}{{k}_{2}}{{\mu }_{1}}{{\mu }_{2}}{{\left| {{x}_{1}} \right|}^{-\frac{1}{2}}}+{{k}_{2}}\mu _{2}^{2}.
\label{eq_polynomial}
\end{equation}

By Vieta's formula, we can derive
\begin{eqnarray}
{{\lambda }_{1}}({{x}_{1}})+{{\lambda }_{2}}({{x}_{1}})&=&-{{k}_{1}}\left({{\mu }_{1}}{{\left| {{x}_{1}} \right|}^{-\frac{1}{2}}}+{{\mu }_{2}}\right), \nonumber\\
   {{\lambda }_{1}}({{x}_{1}})\cdot {{\lambda }_{2}}({{x}_{1}})&=&{{k}_{2}}\left(\tfrac{1}{2}\mu _{1}^{2}{{\left| {{x}_{1}} \right|}^{-1}}+\tfrac{3}{2}{{\mu }_{1}}{{\mu }_{2}}{{\left| {{x}_{1}} \right|}^{-\frac{1}{2}}}+\mu _{2}^{2}\right).
\label{eq_eigenvalue_vieta}
\end{eqnarray}

The design of the discrete version of GSTA relies on the discrete plant
dynamics (\ref{zoh_dynamics}), which can be rewritten as
\begin{equation}
\left[ \begin{matrix}
   {{x}_{1,k+1}}  \\
   {{\varphi }_{k+1}}  \\
\end{matrix} \right]=\left[ \begin{matrix}
   1 & h  \\
   0 & 1  \\
\end{matrix} \right]\left[ \begin{matrix}
   {{x}_{1,k}}  \\
   {{\varphi }_{k}}  \\
\end{matrix} \right]+\left[ \begin{matrix}
   h  \\
   0  \\
\end{matrix} \right]{{u}_{k}}+\left[ \begin{matrix}
   {{h}^{2}}{{\Delta }_{1,k}}/2  \\
   h{{\Delta }_{2,k}}  \\
\end{matrix} \right].
\label{eq_discrete_plant_matrix}
\end{equation}

Similar to Reference \citenum{koch2019discrete}, a general discrete-time controller of the form
\begin{eqnarray}
{{u}_{k}}&=&\frac{1}{h}\left[ -{{x}_{1,k}}+{{{\tilde{u}}}_{1,k}}(h,{{x}_{1,k}}){{x}_{1,k}} \right]+{{\nu }_{k+1}}, \nonumber\\
{{\nu }_{k+1}}&=&{{\nu }_{k}}+{{{\tilde{u}}}_{2,k}}(h,{{x}_{1,k}}){{x}_{1,k}},
\label{eq_general_dc}
\end{eqnarray}
\noindent is used. It should noted that the general controller (\ref{eq_general_dc}) used here is different from the one in Reference \citenum{koch2019discrete}. The original ${{\nu}_{k}}$ in the first line of the general controller is replaced by ${{\nu}_{k+1}}$ which eliminates the usage of the previous system state. This modification can result in some performance improvement.

Substituting (\ref{eq_general_dc}) into (\ref{eq_discrete_plant_matrix}) yields the discrete-time closed-loop dynamics
\begin{equation}
\boldsymbol{x}_{k+1}={{\boldsymbol{M}}_{d}}({{x}_{1,k}}){{\boldsymbol{x}}_{k}}+h{{\boldsymbol{d}}_{k}},
\label{eq_discrete_sys_matrix}
\end{equation}
\noindent with \({{\boldsymbol{x}}_{k}}={{[{{x}_{1,k}},{{x}_{2,k}}]}^{T}}\), ${{\boldsymbol{d}}_{k}}={[{h{{\Delta }_{1,k}}/2,{{\Delta }_{2,k}}]}^{T}}$, and
 
\begin{equation}
\boldsymbol{M}_{d}({{x}_{1,k}})=\left[ \begin{matrix}
   {{{\tilde{u}}}_{1,k}}(h,{{x}_{1,k}}) & h  \\
   {{{\tilde{u}}}_{2,k}}(h,{{x}_{1,k}}) & 1  \\
\end{matrix} \right].
\label{eq_M_d}
\end{equation}

The function \({{\tilde{u}}_{1,k}}(h,{{x}_{1,k}})\) can be designed by the eigenvalue mapping-based method. Assuming ${{\boldsymbol{M}}_{d}}({{x}_{1,k}})$ has real or a pair of conjugate complex eigenvalues noted as ${{q}_{1}}(h,{{x}_{1,k}})$ and ${{q}_{2}}(h,{{x}_{1,k}})$. Then we can easily deduce
\begin{eqnarray}
{{{\tilde{u}}}_{1,k}}(h,{{x}_{1,k}})&=&{{q}_{1}}(h,{{x}_{1,k}})+{{q}_{2}}(h,{{x}_{1,k}})-1, \nonumber\\
   {{{\tilde{u}}}_{2,k}}(h,{{x}_{1,k}})&=&\frac{1}{h}\left[ {{{\tilde{u}}}_{1,k}}(h,{{x}_{1,k}})-{{q}_{1}}(h,{{x}_{1,k}}){{q}_{2}}(h,{{x}_{1,k}}) \right].
\label{eq_u_k_relate_q_k}
\end{eqnarray}

Now the discretization of GSTA becomes the problem of finding appropriated eigenvalues of ${{\boldsymbol{M}}_{d}}({{x}_{1,k}})$. Since we have known the eigenvalue of $\boldsymbol{M}({{x}_{1}})$, It's very natural to think about how to establish the relationship between the two matrix's eigenvalues. We start with the simplest explicit Euler method and find the corresponding eigenvalues mapping relationship. The closed-loop system (\ref{closed_loop_explicit_euler}) resulting from the explicit Euler method can be rewritten as
\begin{equation}
\boldsymbol{x}_{k+1}=[\boldsymbol{I}+h\boldsymbol{M}({{x}_{1,k}})]{{\boldsymbol{x}}_{k}}+h{{\boldsymbol{d}}_{k}}.
\end{equation}

The eigenvalues \({{q}_{n}}(h,{{x}_{1,k}})\) of \([\boldsymbol{I}+h\boldsymbol{M}({{x}_{1,k}})]\) are located at \(1+h{{\lambda }_{i}}({{x}_{1,k}})\) for \({{x}_{1,k}}\ne 0\) and \(\underset{{{x}_{1,k}}\to 0}{\mathop{\lim }}\,\left| {{q}_{n}}(h,{{x}_{1,k}}) \right|=\infty \). Thus, the explicit eigenvalue mapping is
\begin{equation}
q_{n}(h,{{x}_{1,k}})=
\begin{cases}
{1+h{{\lambda}_{n}}({{x}_{1,k}})}& \text{ ${{x}_{1,k}}\ne 0$ } \\
0& \text{ ${{x}_{1,k}}=0$ }
\end{cases}.
\label{eq_relu_mapping}
\end{equation}

As shown in Reference \citenum{koch2019discrete}, there are some other eigenvalue mappings, like the implicit mapping and the matching approach. The implicit mapping is
\begin{equation}
q_{n}(h,{{x}_{1,k}})=
\begin{cases}
\frac{1}{1-h{{\lambda }_{n}}({{x}_{1,k}})}& \text{ $\quad {{x}_{1,k}}\ne 0$ } \\
0& \text{ $ \quad {{x}_{1,k}}=0 $ }
\end{cases}.
\label{eq_semi_implicit_mapping}
\end{equation}

It's worth noting that implicit mapping (\ref{eq_semi_implicit_mapping}) may run into a singularity problem when ${1-h{{\lambda }_{n}}({{x}_{1,k}})=0}$.
Fortunately, According to equation (\ref{eq_eigenvalue_vieta}), we can easily deduce 
${{\lambda }_{n}}<0$, then ${1-h{{\lambda }_{n}}({{x}_{1,k}})>0}$. Therefore, this case will never happen.

For the matching approach, its eigenvalue mapping is
\begin{equation}
q_{n}(h,{{x}_{1,k}})=
\begin{cases}
{{e}^{h{{\lambda }_{n}}({{x}_{1,k}})}}& \text{ $\quad {{x}_{1,k}}\ne 0$ } \\
0& \text{ $ \quad {{x}_{1,k}}=0 $ }
\end{cases}
\label{eq_matching_mapping}
\end{equation}

Noted that roots of the polynomial (\ref{eq_polynomial}) may have imaginary roots, then we can not get $q_{n}(h,{{x}_{1,k}})$ directly. For the standard super twisting algorithm, we can simply set $k_{1}^{2} \ge 2{{k}_{2}}$ to avoid this case. This becomes a little complicated when extending it to GSTA. For ${{x}_{1,k}}\ne 0$, it can be seen that the controller (\ref{eq_general_dc}) only contains the sum and product of the EMF value. This means specific EMF values are not required. For the explicit mapping and the implicit mapping, we find the sum and product of the EMF value can be easily rewritten as a mixture of the sum and product of the eigenvalues which is already known according to equation (\ref{eq_eigenvalue_vieta}). Therefore, the controller discretized by these two methods is totally independent of the imaginary roots and has no need to calculate the specific roots. However, this is impossible for the matching approach. Fortunately, we find that the imaginary root is allowable when using Euler’s formula to calculate the control input indirectly. According to polynomial (\ref{eq_polynomial}), ${{\lambda }_{1,k}}$ and ${{\lambda }_{2,k}}$ can be rewritten as ${{\lambda }_{1,k}}={{\lambda }_{r,k}}+{{\lambda }_{i,k}}i$ and ${{\lambda }_{2,k}}={{\lambda }_{r,k}}-{{\lambda }_{i,k}}i$. By Euler's formula, ${{e}^{h{{\lambda }_{1,k}}}}$ can be represented as ${{e}^{h{{\lambda }_{r,k}}}}\left[ \cos (h{{\lambda }_{i,k}})+i\sin (h{{\lambda }_{i,k}}) \right]$, ${{e}^{h{{\lambda }_{2,k}}}}$ can be represented as ${{e}^{h{{\lambda }_{r,k}}}}\left[ \cos (h{{\lambda }_{i,k}})-i\sin (h{{\lambda }_{i,k}}) \right]$. Then we can deduce that ${{q}_{1}}(h,{{x}_{1,k}})+{{q}_{1}}(h,{{x}_{1,k}})=2{{e}^{{{\lambda }_{r,k}}}}\cos (h{{\lambda }_{i,k}})$ and ${{q}_{1}}(h,{{x}_{1,k}}){{q}_{1}}(h,{{x}_{1,k}})={{e}^{2{{\lambda }_{r,k}}}}$. Thus, the matching approach allows the existence of the imaginary root.




\begin{figure}[tb]
\centering
\subfloat[\label{0a}]{
\includegraphics[scale=1]{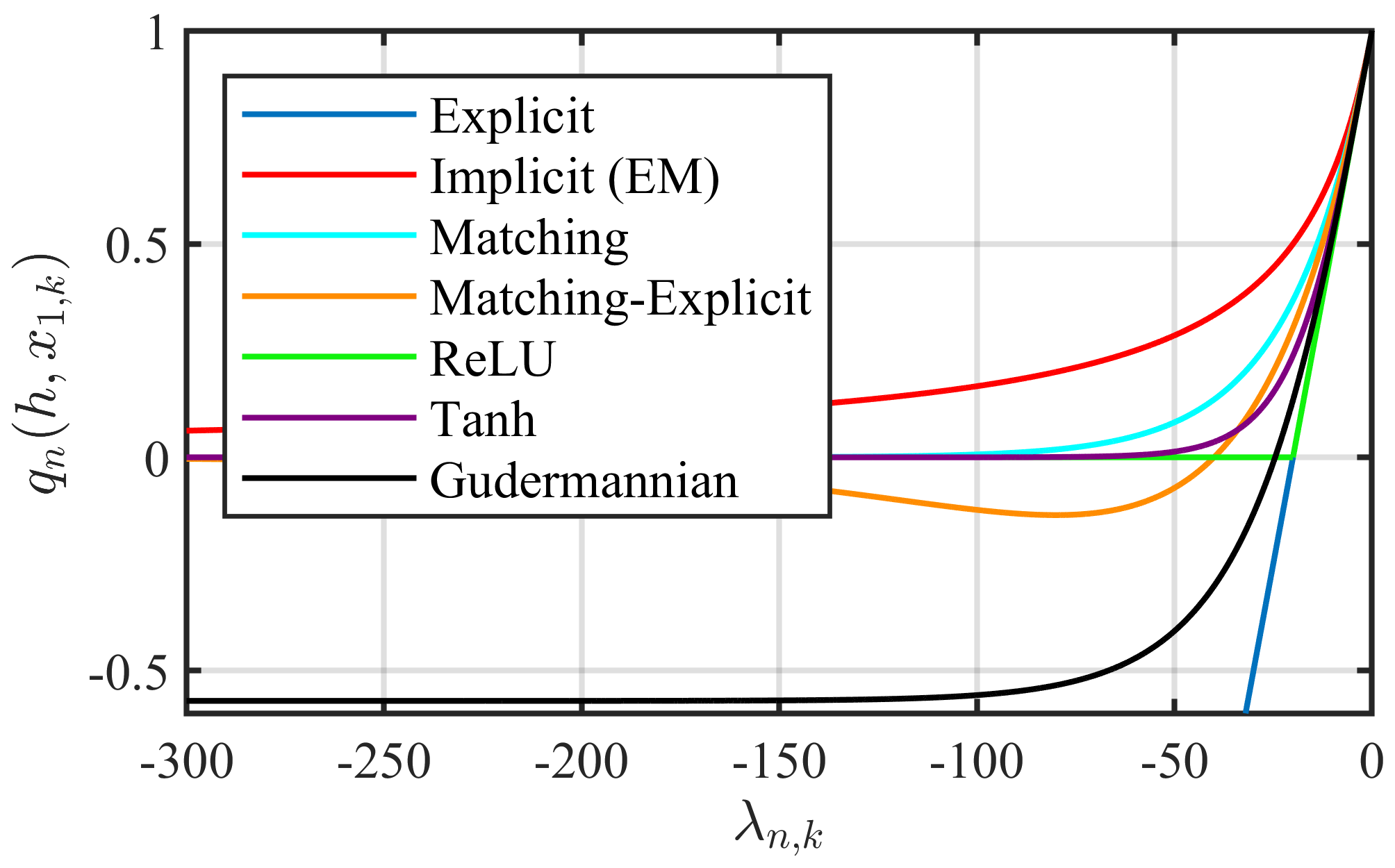}}
\subfloat[\label{0b}]{    
\includegraphics[scale=1]{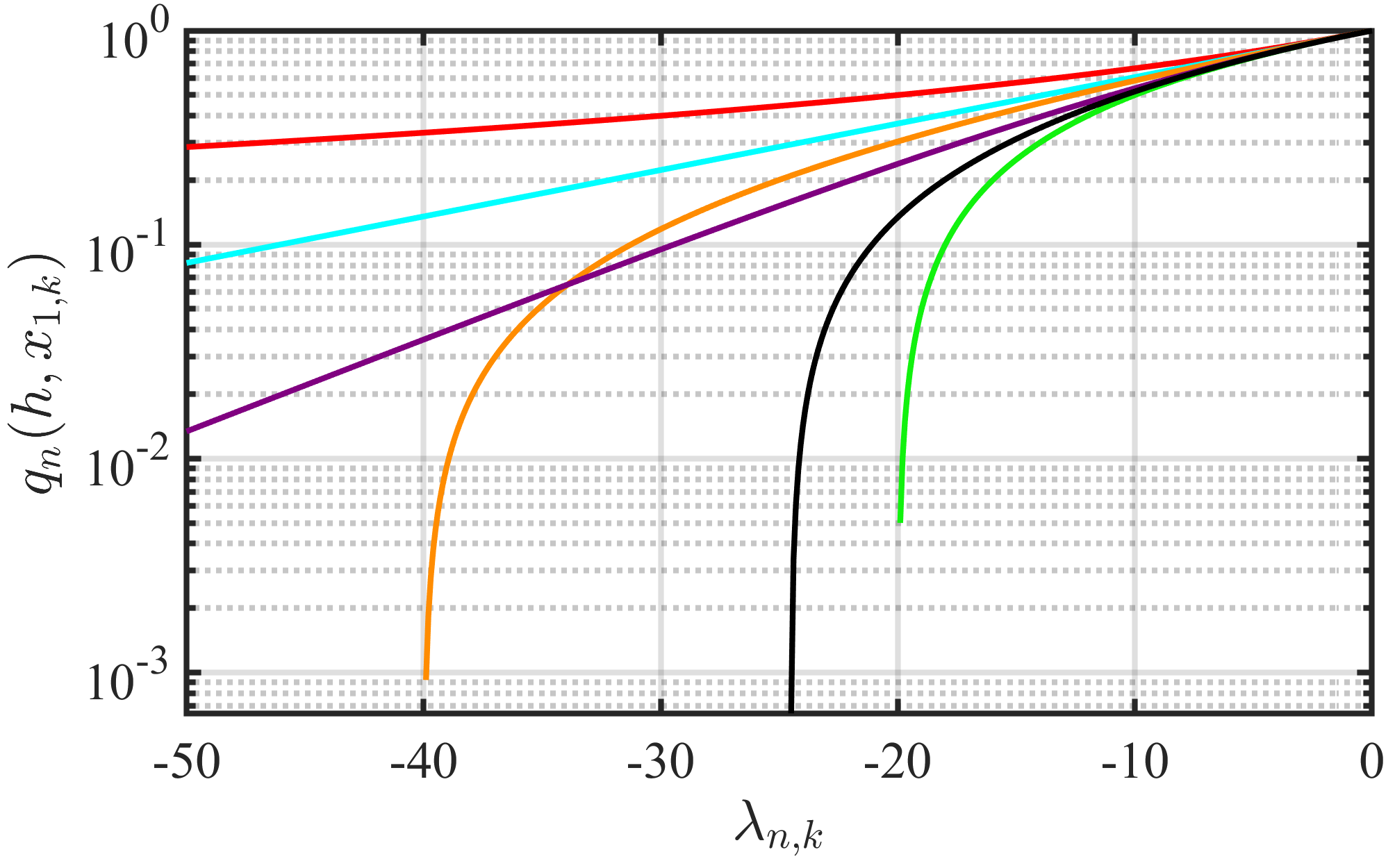}}
\caption{Curve of eigenvalue functions. The sampling time $h$ is 0.05s}
\label{fig:em_function}
\end{figure}

Besides the above-mentioned eigenvalue mappings, we also want to explore new eigenvalue mappings to achieve better control performance. One simple idea is to plot all eigenvalue mapping functions, as shown in Figure \ref{fig:em_function}, and observe their property. We find that all eigenvalue mapping functions start from 1 and decrease as ${\lambda}_{n}({{x}_{1,k}})$ decreases. Besides, it also satisfies $q_{n}(h,{{x}_{1,k}})=0$ for ${{x}_{1,k}}=0$ which is used to avoid singularity. To ensure the controller (\ref{eq_general_dc}) is indeed a discrete-time implementation of GSTA, the solutions of the discrete-time system (\ref{eq_discrete_sys_matrix}) should approximate the solutions of the continuous-time system (\ref{eq_continue_matrix}) in any compact set not including points ${x}_{1,k}=0$. Compute the limit $\underset{h\to 0}{\mathop{\lim }}\,\frac{1}{h}({{\boldsymbol{x}}_{k+1}}-{{\boldsymbol{x}}_{k}})=\underset{h\to 0}{\mathop{\lim }}\,\frac{1}{h}({{\boldsymbol{M}}_{d}}({{x}_{1,k}}){{\boldsymbol{x}}_{k}}-{{\boldsymbol{x}}_{k}})$ with (\ref{eq_M_d}), (\ref{eq_u_k_relate_q_k}) and for any ${{x}_{1,k}}\ne0$ yields

\begin{eqnarray}
  &&\underset{h\to 0}{\mathop{\lim }}\,\frac{1}{h}({{\boldsymbol{x}}_{k+1}}-{{\boldsymbol{x}}_{k}}) = \underset{h\to 0}{\mathop{\lim }}\,\left[ \begin{array}{*{35}{l}}
   \frac{1}{h}({{q}_{1}}+{{q}_{2}}-2){{x}_{1,k}}+{{x}_{2,k}}+h{{\Delta }_{1,k}}/2  \\
   \frac{1}{{{h}^{2}}}({{q}_{1}}+{{q}_{2}}-{{q}_{1}}{{q}_{2}}-1){{x}_{1,k}}+{{\Delta }_{2,k}}  \\
\end{array} \right] = \left[ \begin{array}{*{35}{c}}
-{{k}_{1}}\left({{\mu }_{1}}{{\left| {{x}_{1}} \right|}^{-\frac{1}{2}}}+{{\mu }_{2}}\right){{x}_{1}}+{{x}_{2}}  \\
-{{k}_{2}}\left(\tfrac{1}{2}\mu _{1}^{2}{{\left| {{x}_{1}} \right|}^{-1}}+\tfrac{3}{2}{{\mu }_{1}}{{\mu }_{2}}{{\left| {{x}_{1}} \right|}^{-\frac{1}{2}}}+\mu _{2}^{2}\right){{x}_{1}}+\Delta (t)  \\
\end{array} \right]. \nonumber \\
\label{eq_approximate}
\end{eqnarray}


To ensure equation (\ref{eq_approximate}) holds, the new EMF is supposed to satisfy the following constraints:

\begin{equation}
\underset{h\to 0}{\mathop{\lim }}\,\frac{1}{h}({{q}_{1}}+{{q}_{2}}-2)=
\begin{cases}
{{\lambda }_{1}}+{{\lambda }_{2}} & \textit{real roots} \\
2{{\lambda }_{r}} & \textit{others}
\end{cases},
\label{eq_approximate_1}
\end{equation}

\begin{equation}
\underset{h\to 0}{\mathop{\lim }}\,\frac{1}{{{h}^{2}}}({{q}_{1}}+{{q}_{2}}-{{q}_{1}}{{q}_{2}}-1)=
\begin{cases}
-{{\lambda }_{1}}{{\lambda }_{2}} & \textit{real roots} \\
-\lambda _{r}^{2}-\lambda _{i}^{2} & \textit{others}
\end{cases}.
\label{eq_approximate_2}
\end{equation}

One simple idea to find new EMFs is to combine different EMFs. For example, the hybrid of the matching approach and explicit mapping is
\begin{equation}
q_{n}(h,{{x}_{1,k}})=
\begin{cases}
{{e}^{\frac{1}{2}h{{\lambda }_{n}}({{x}_{1,k}})}}\left( 1+\frac{1}{2}h{{\lambda}_{n}}({{x}_{1,k}}) \right)& \text{ $\quad {{x}_{1,k}}\ne 0$ } \\
0& \text{ $ \quad {{x}_{1,k}}=0 $ }
\end{cases}.
\label{eq_matching_explicit_mapping}
\end{equation}
From Figure \ref{fig:em_function}, we find that the matching-explicit EMF approximates 0 from the negative constant which is different from others. But it satisfies the basic constraint of EMFs. In APPENDIX A, we prove that the discrete-time controller using the matching-explicit mapping (\ref{eq_matching_explicit_mapping}) does approximate the continuous-time controller when the sampling time approaches 0. It's worth noting that the hybrid of the EMFs is not just a simple multiplication. The coefficient changes accordingly.

In Figure \ref{fig:em_function}, we observe that only the explicit EMF becomes negative infinity as ${{\lambda }_{n,k}}$ tends to negative infinity. From equation (\ref{eq_eigenvalue_vieta}), we can derive that ${{\lambda }_{n,k}}$ tends to negative infinity when ${{x}_{1,k}}$ converges to origin. According to equations (\ref{eq_general_dc}) and (\ref{eq_u_k_relate_q_k}), the control input by explicit mapping may not converge to zero as the state converges to zero. This explains the discretization chattering that appears in the explicit discretized GSTA. To avoid this problem, the EMF should not become negative infinity as ${{x}_{1,k}}$ is steered to zero. Inspired by the ReLU function that is often used as an activation function in the machine learning domain, we proposed a new eigenvalue mapping
\begin{equation}
q_{i}(h,{{x}_{1,k}})=
\begin{cases}
{1+h{\lambda_{n}}({x}_{1,k})}& \text{ $ {{\lambda }_{n,k(r,k)}}({{x}_{1,k}})>-\frac{1}{h},  {{x}_{1,k}}\ne 0$ } \\
0& \text{ $else$ }
\end{cases}.
\label{eq_relu_mapping}
\end{equation}
The proposed ReLU mapping function can be seen as a truncated explicit eigenvalue mapping function. We will call it the ReLU mapping due to its similarity to the ReLU function. When the sampling time approaches 0, the ReLU mapping is equivalent to the known explicit mapping. Thus, the discrete-time controller using the ReLU EMF (\ref{eq_relu_mapping}) also approximates the continuous-time controller.

Inspired by the hyperbolic tangent function which is also often used as an activation function in the machine learning domain, we proposed a new EMF
\begin{equation}
q_{n}(h,{{x}_{1,k}})=
\begin{cases}
\frac{\text{2}{{e}^{h{{\lambda }_{n,k}}({{x}_{1,k}})}}}{{{e}^{h{{\lambda }_{n,k}}({x}_{1,k})}}\text{+}{{e}^{-h{{\lambda }_{n}}({x}_{1,k})}}}& \text{ $\quad {{x}_{1,k}}\ne 0$ } \\
0& \text{ $ \quad {{x}_{1,k}}=0 $ }
\end{cases}.
\label{eq_tanh_mapping}
\end{equation}
As shown in Figure \ref{fig:em_function}(b), the hyperbolic tangent function's convergence rate is much faster than the Matching EMF and Implicit EMF and is also faster than the Matching-Explicit EMF when the EMF is positive. This means that the hyperbolic tangent (Tanh) mapping may provide better control performance. In APPENDIX B, it has been proven that the Gudermannian discretized GSTA is the discrete-time version of the GSTA.

Inspired by the Gudermannian function, we proposed another new EMF
\begin{equation}
q_{n}(h,{{x}_{1,k}})=
\begin{cases}
\arctan \left( \sinh \left( h{{\lambda }_{n}}\left( {{x}_{1,k}} \right) \right) \right)+1& \text{ $\quad {{x}_{1,k}}\ne 0$ } \\
0& \text{ $ \quad {{x}_{1,k}}=0 $ }
\end{cases}.
\label{eq_good_mapping}
\end{equation}
As shown in Figure \ref{fig:em_function}(a), the Gudermannian EMF's convergence rate is the fastest except for ReLU EMF when EMF remains positive. The Gudermannian EMF will converge to a negative constant instead of zero if ${{\lambda }_{n}}$ tends to negative infinity. In APPENDIX C, it has been proven that the Gudermannian discretized GSTA is the discrete-time version of the GSTA.

\section{SIMULATION EXAMPLES}
In this section, the effectiveness of the proposed discretization schemes is verified by numerical simulations. All simulations are implemented in MATLAB/Simulink. In this section, we use ${{\Sigma }_{*}}$ to denote the simulated closed-loop system consisting of the simulated continuous-time model and the corresponding discrete-time controller. The simulation files are available at \url{https://github.com/HMX2013/GSTA-Discretization}

\begin{figure}[htbp]
\centering
\subfloat[\label{0a}][$\Lambda=4$]{
    \includegraphics[scale=1]{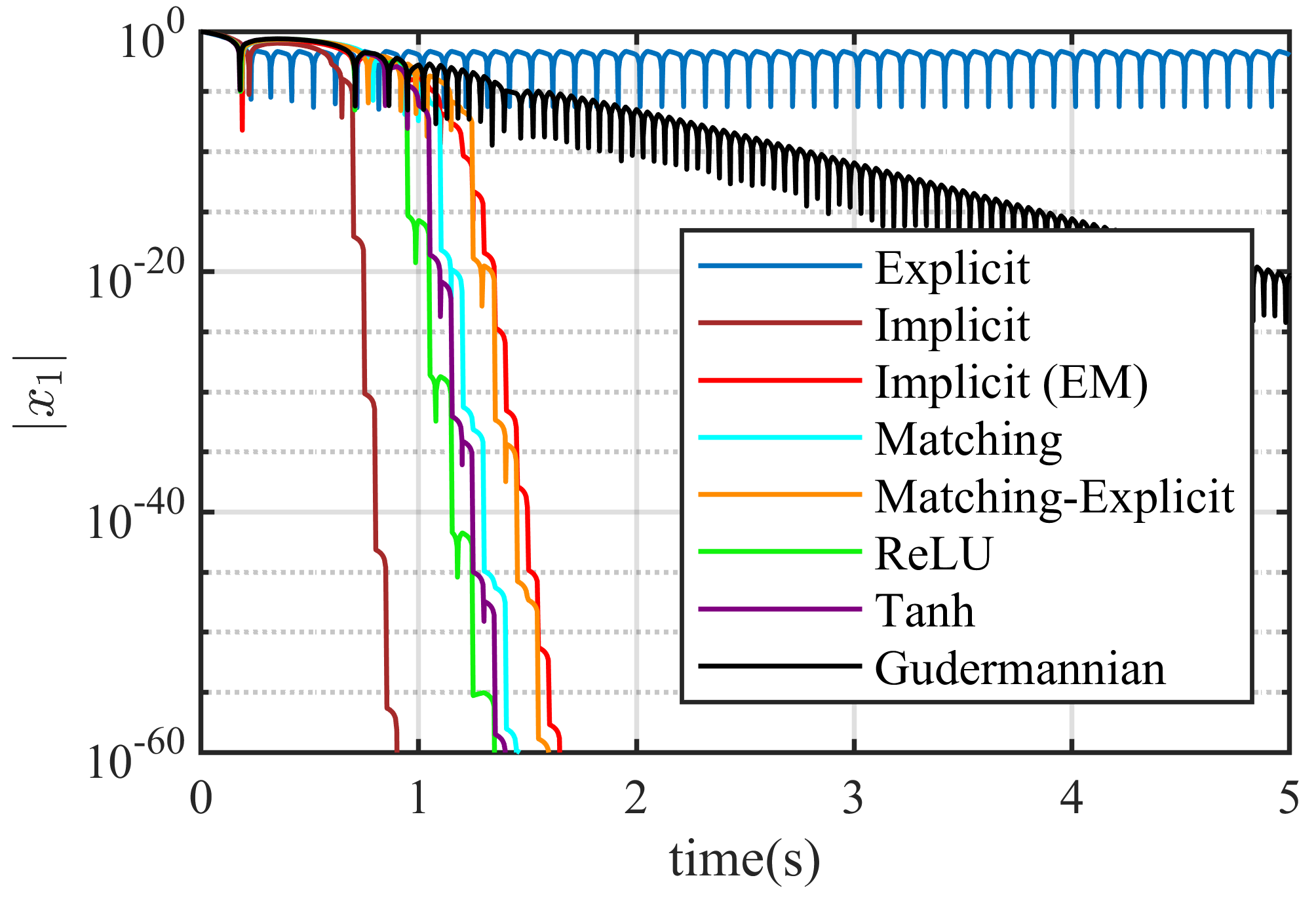}}
\subfloat[\label{0b}][$\Lambda=40$]{    
    \includegraphics[scale=1]{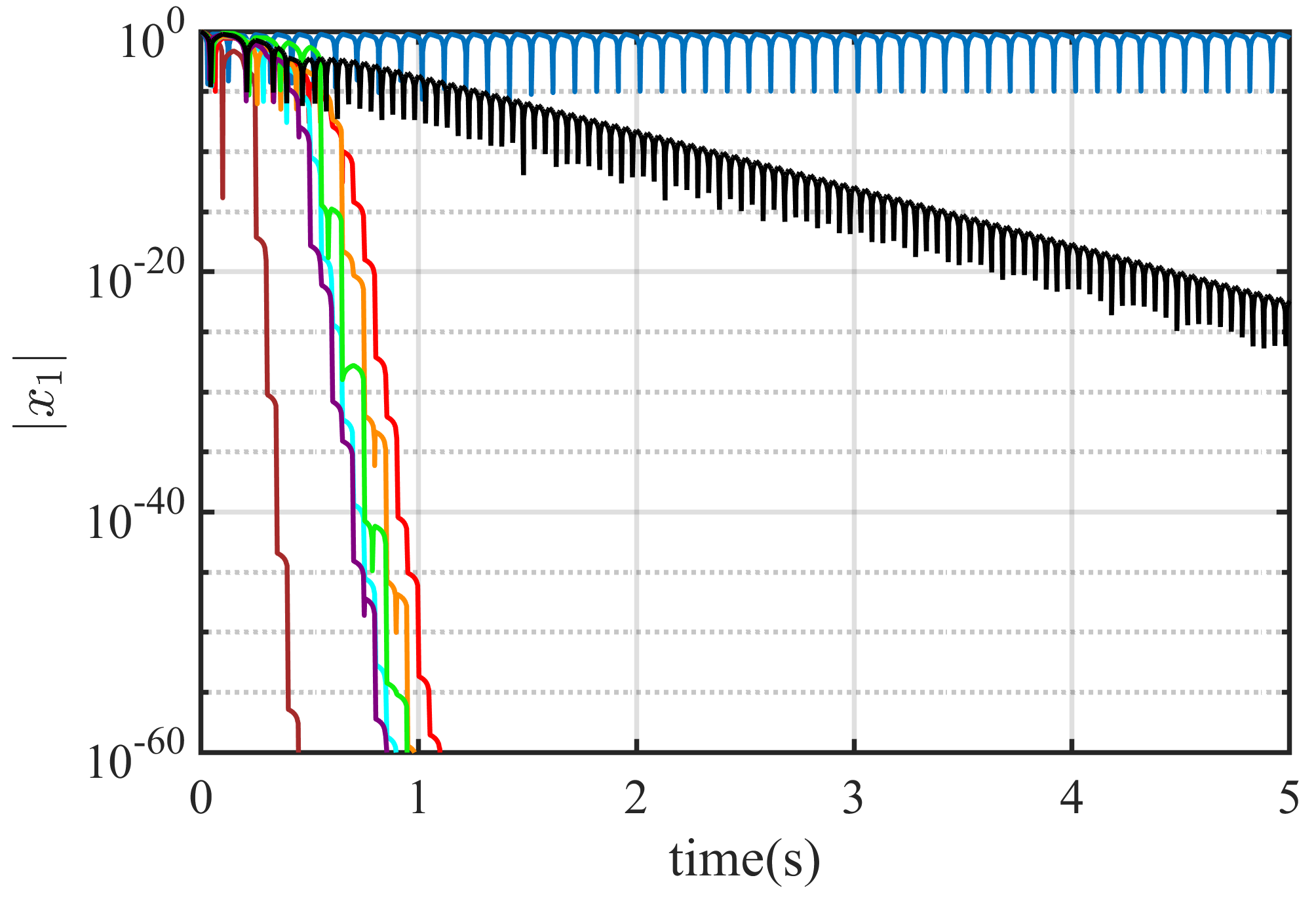}}
\caption{Comparison of the convergence rate of the state variable $\left| {{x}_{1}} \right|$. The parameters are ${{k}_{1}}=1.5\sqrt{\Lambda}$, ${{k}_{2}}=2.2\Lambda$, ${{\mu }_{1}}=1$, ${{\mu}_{2}}=1$ and $h=0.05$.}
\label{fig:undisturb_effect}
\end{figure}

\subsection{Simulations without Disturbance in Time-Domain}
\label{sim_undisturb}
The first group simulations were performed without disturbance, ie. $\Delta \equiv 0$. In Figure \ref{fig:undisturb_effect}(a), parameters were chosen as ${{k}_{1}}=1.5\sqrt{\Lambda}$, ${{k}_{2}}=2.2\Lambda$, $\Lambda=4$, ${{\mu }_{1}}=1$, ${{\mu}_{2}}=1$, $h=0.05$ and $x_{1}(0)=1$. In Figure \ref{fig:undisturb_effect}(b), $\Lambda$ was increased to 40 and other parameters remained unchanged. The system state $x_{1}$ is depicted in absolute values and scaled logarithmically to emphasize the differences. The results show that $\Sigma_{explicit}$ is not exact and shows obvious discretization chattering in the steady state. All other systems converge to zero without discretization-chattering. However, ${{\sum }_{\rm Gudermannian}}$ converges much slower than other systems. This is caused by the constant negative EMF when state $x_1$ is close to the origin. In contrast,${{\sum }_{\rm Implicit}}$ provides the fastest convergence speed which shows its superiority in the unperturbed case. For the positive and smooth EMF, it can be observed the system convergence speed presented in Figure \ref{fig:undisturb_effect} is consistent with the EMF's convergence speed as shown in Figure \ref{fig:em_function}. However, the other types of EMFs don't follow this pattern. For example, ${{\sum }_{\rm Tanh}}$ converges faster than ${{\sum }_{\rm ReLU}}$ when increasing $\Lambda$ from 4 to 40. All systems except ${{\sum }_{\rm Explicit}}$ show faster convergence speed when increasing the control gains. The simulation results indicate that the system convergence speed is strongly linked to the convergence speed of the corresponding EMF.

\begin{figure*}
\centerline{\includegraphics[scale=1.0]{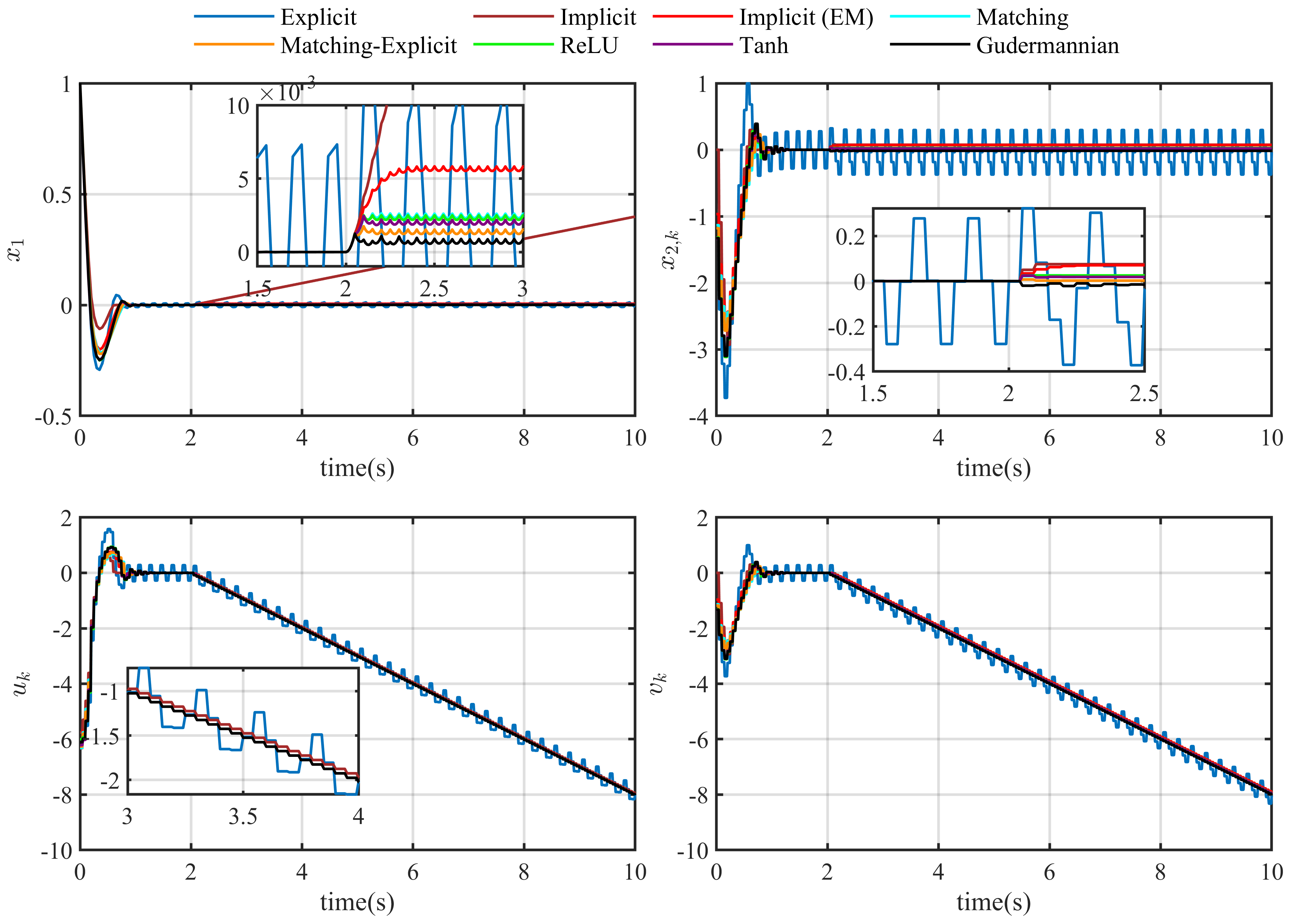}}
\caption{The system states $x_{1,k}$, $x_{2,k}$, the control input $u_{k}$ and the variable $\nu_{k}$ of the
system (4) in the time domain in the unbounded disturbance case.
\label{fig:disturb_effect_1}}
\end{figure*}

\subsection{Simulations with Disturbance in Time-Domain}
\label{sim_disturb}
\begin{figure*}
\centerline{\includegraphics[scale=1.0]{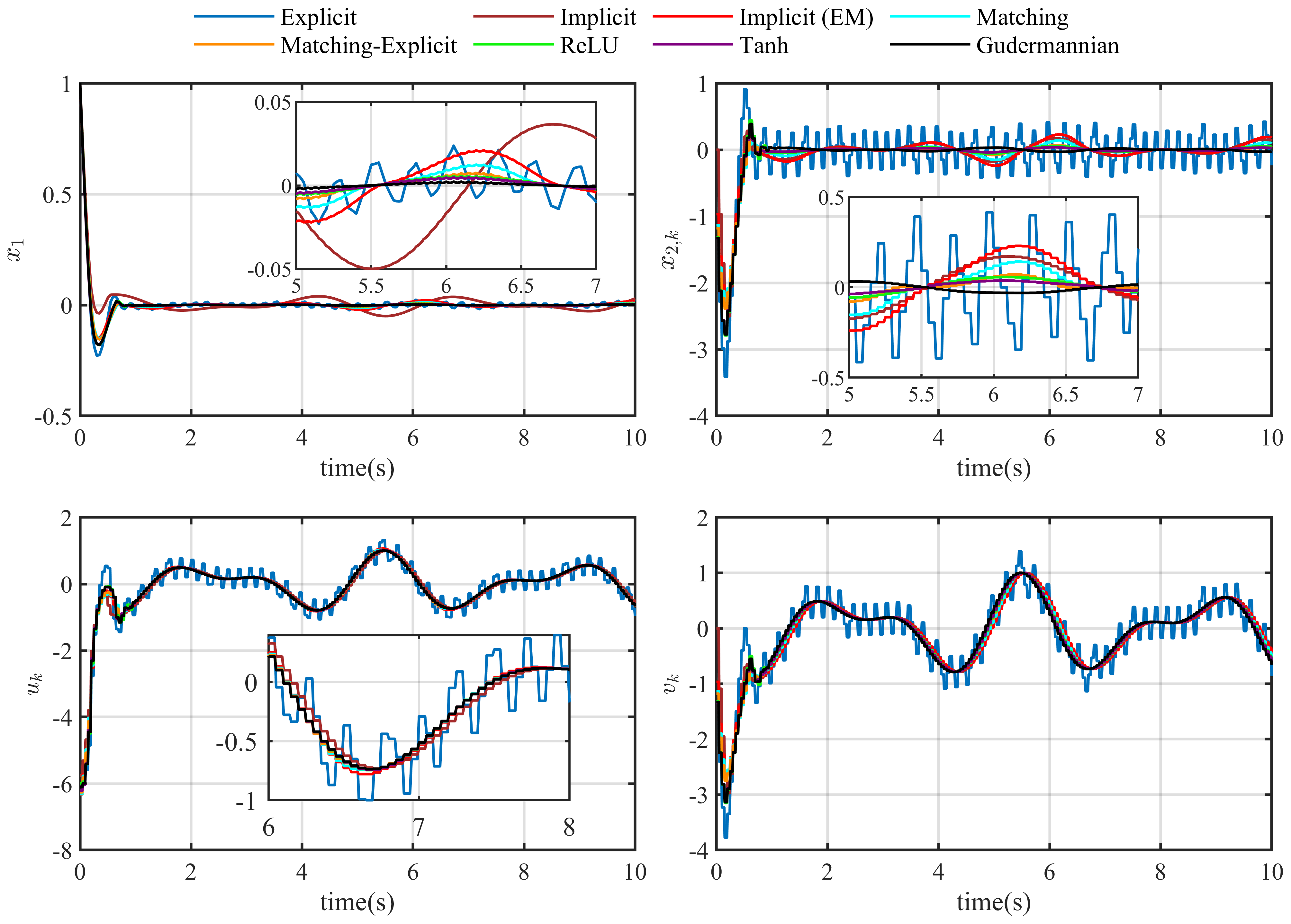}}
\caption{The system states $x_{1,k}$, $x_{2,k}$, the control input $u_{k}$ and the variable $\nu_{k}$ of the
system (4) in the time domain in the bounded disturbance case.
\label{fig:disturb_effect_2}}
\end{figure*}

Firstly, we consider the constant disturbance $\Delta =0,\forall t<2$ and $\Delta =1,\forall t\ge 2$.
In this case, the perturbation $\varphi (t)$ will grow unbounded. The other parameters setting is the same as section \ref{sim_undisturb} ($\Lambda=4$). The simulation results are presented in Figure \ref{fig:disturb_effect_1}. The results clearly show that only the $\Sigma_{\rm Implicit}$ is not capable of rejecting unbounded disturbance $\varphi (t)$ which is generated by constant parts of the disturbance $\Delta(t)$. 
This means the implicit Euler method reduces the class of disturbances $\varphi(t)$ that can be handled by the discrete-time controller compared to the continuous-time GSTA. All other controllers result in the state $x_1$ converging close to the neighborhood of origin. However, $\Sigma_{\rm Explicit}$ exists severe discretization chattering as expected. The enlarged view in Figure \ref{fig:disturb_effect_1} shows that our proposed methods, especially Gudermannian EMF provide smaller steady-state error than the original ones. Then we consider a bounded disturbance that has been widely used in the literature,\cite{koch2019discrete, andritsch2023modified, xiong2021discrete} i.e., $\Delta(t) =1.2cos(2t)+0.4\sqrt{10}\cos(\sqrt{10}t)$. The simulation results are shown in Figure \ref{fig:disturb_effect_2}. Different from the unbounded disturbance case, $\Sigma_{\rm Implicit}$ can converge state $x_1$ to the neighborhood of origin. We can also observe the obvious chattering phenomenon for $\Sigma_{\rm Explicit}$. For the other discrete-time controllers, the discretization chattering is eliminated, and the control input is smoother which benefits the practical application. The enlarged view in Figure \ref{fig:disturb_effect_2} once again demonstrates the superiority of our discrete-time version of GSTA.

\subsection{Steady-State Accuracy when Varying the Sampling Time}

\label{sim_samp_time}
\begin{figure}
\centerline{\includegraphics[scale=1]{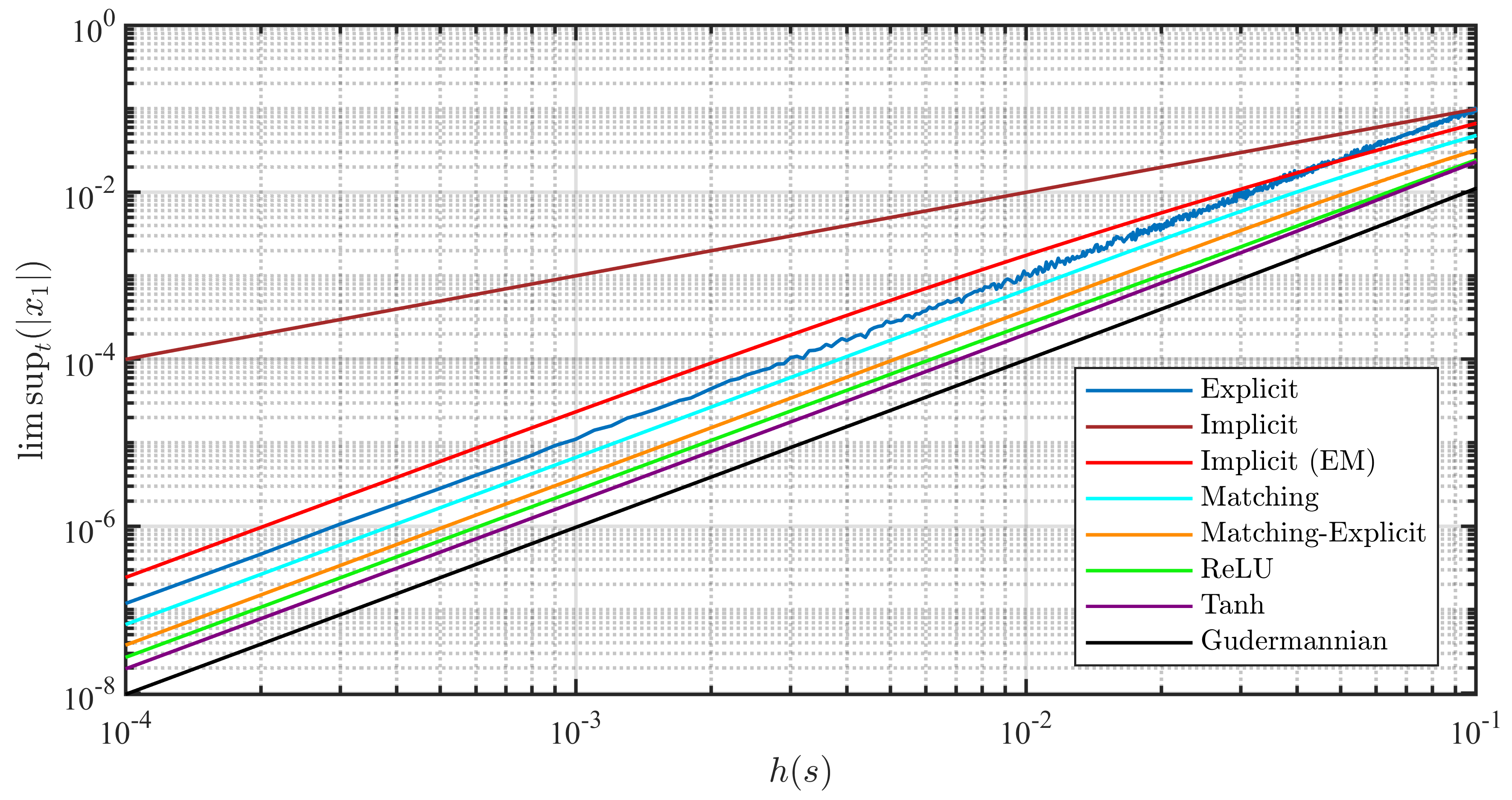}}
\caption{Precision w.r.t. an increase of sampling time.
\label{fig:sampling_effect}}
\end{figure}

In the simulation, all the parameters and the initial states are kept the same as in section \ref{sim_disturb} (Bounded disturbance), while the sampling time variable $h$ is increased from $10^{-4}$ s to $10^{-1}$ s in 1000 steps. The remaining steady-state error (SSE) is used to analyze the accuracy of the closed-loop systems. Let the SSE be defined as ${{e}_{f}}=\lim {{\sup }_{t}}\left| {{x}_{1}} \right|$, with the initial value ${{x}_{1}}(0)=1$. The system simulation time is set to 50s. The inclined lines in Figure \ref{fig:sampling_effect} illustrate the precision level of the system when increasing the sampling time. The Implicit method provides the lowest precision level, which has an obvious gap compared with other methods. The result verifies that the implicit Euler method can only achieve the first-order accuracy of SMC. Compared with the original eigenvalue mapping-based discretization method, the proposed methods provide higher precision. It's worth noting that $\Sigma_{\rm Gudermannian}$ achieves the highest precision and the SSE is very close to $h^{2}$.

\begin{figure}[h]
    \centering
    \subfloat[\label{2a}][Accuracy over $\Delta$]{
        \includegraphics[scale=1.0]{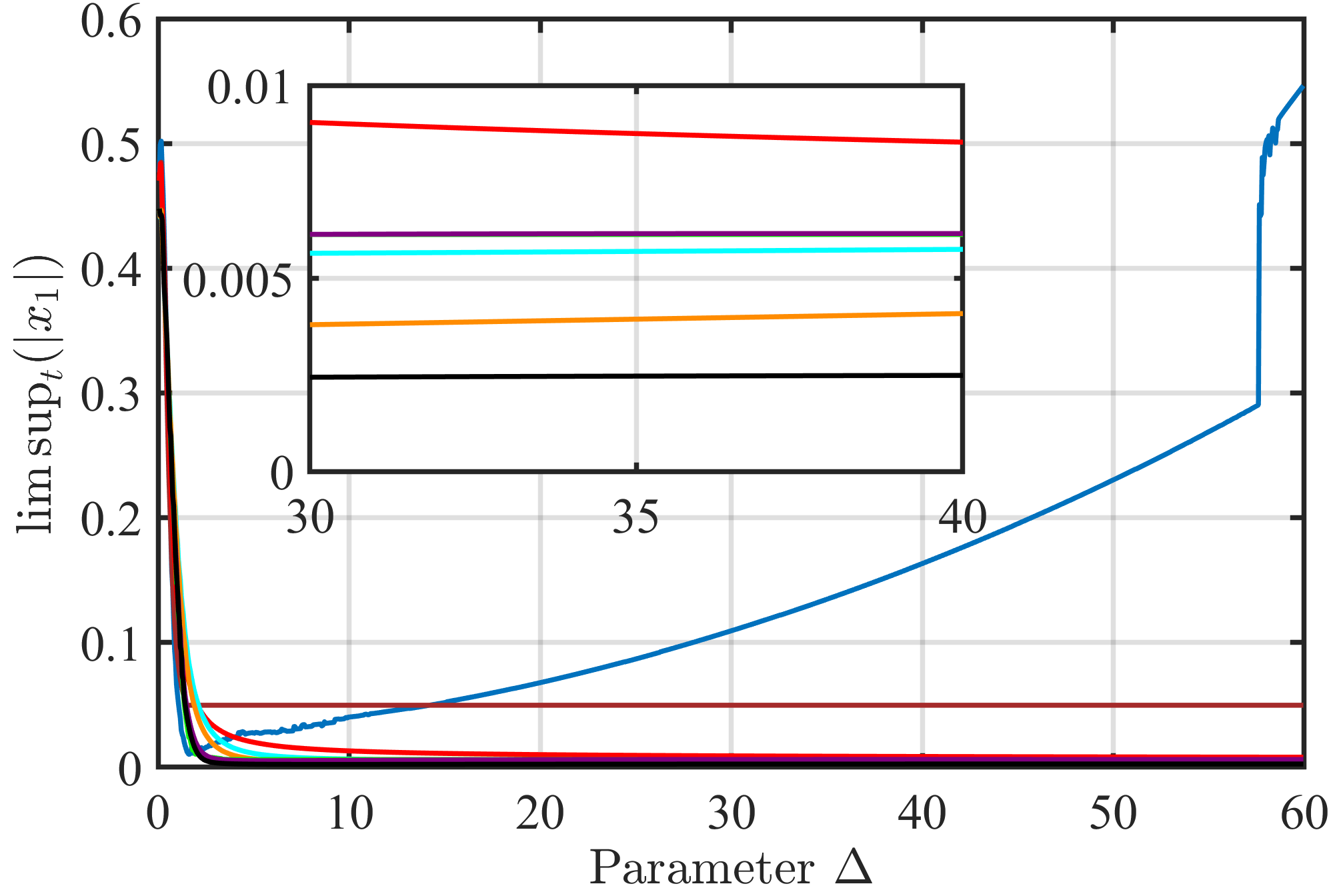}}
    \subfloat[\label{2a}][Accuracy over ${k}_{1}$ with fixed ${k}_{2}=10$]{
        \includegraphics[scale=1.0]{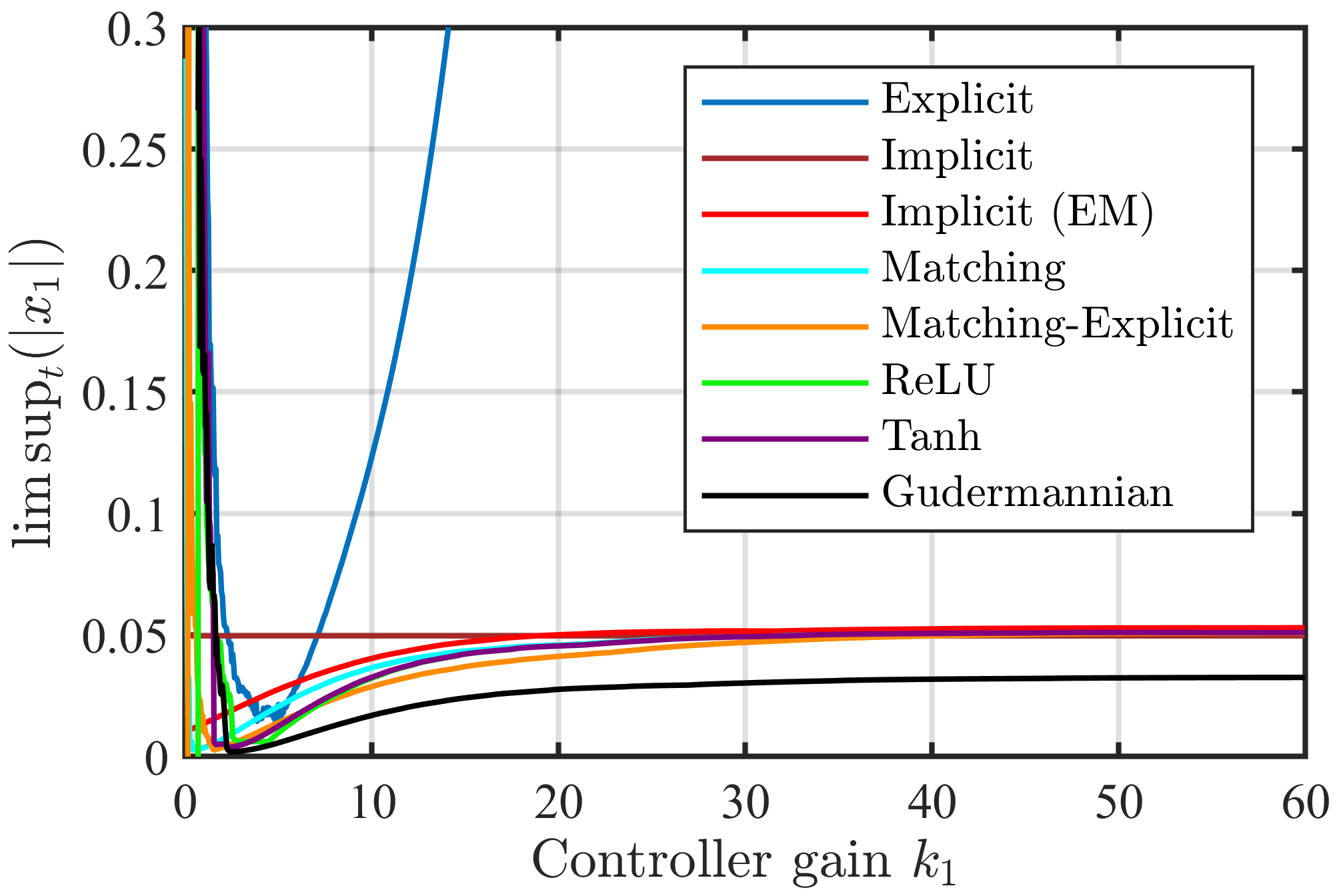}}\hfill
    \subfloat[\label{2b}][Accuracy over ${k}_{2}$ with fixed ${k}_{1}=10$]{    
        \includegraphics[scale=1.0]{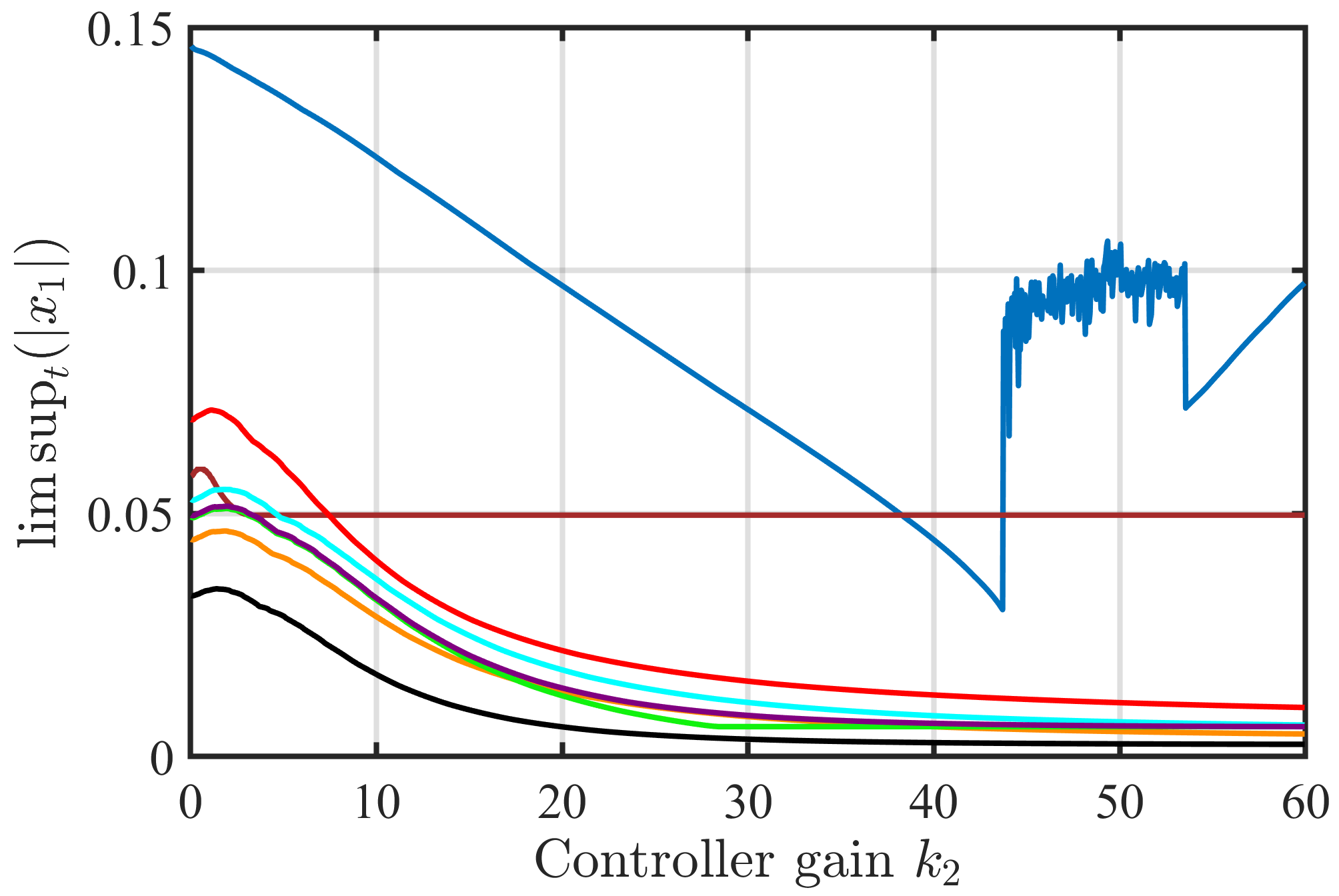}}
    \subfloat[\label{2c}][Accuracy over $\mu_{1}$ with fixed $\mu_{2}=1$]{
        \includegraphics[scale=1.0]{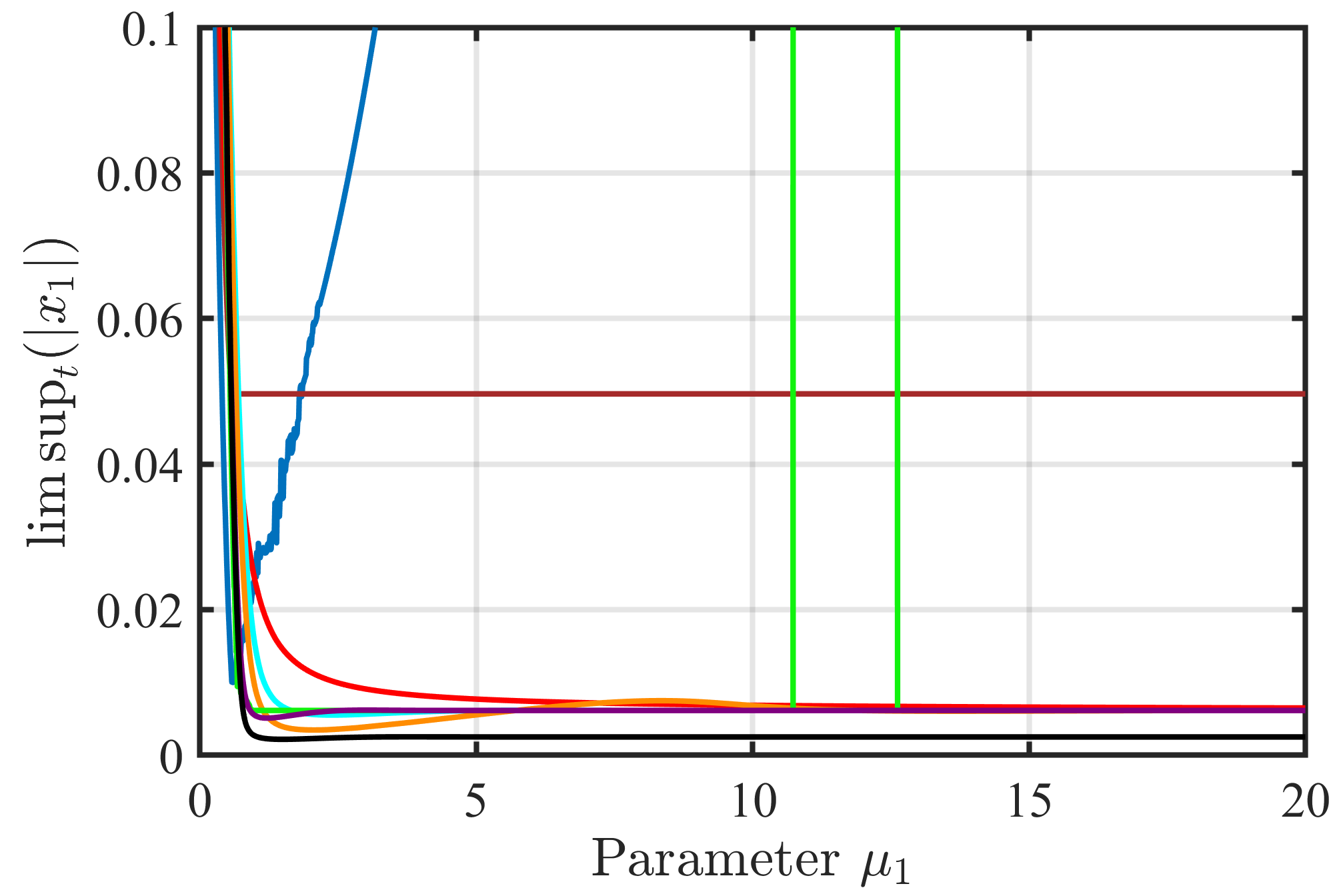}}\hfill
    \subfloat[\label{2d}][Accuracy over $\mu_{2}$ with fixed $\mu_{1}=1$]{
        \includegraphics[scale=1.0]{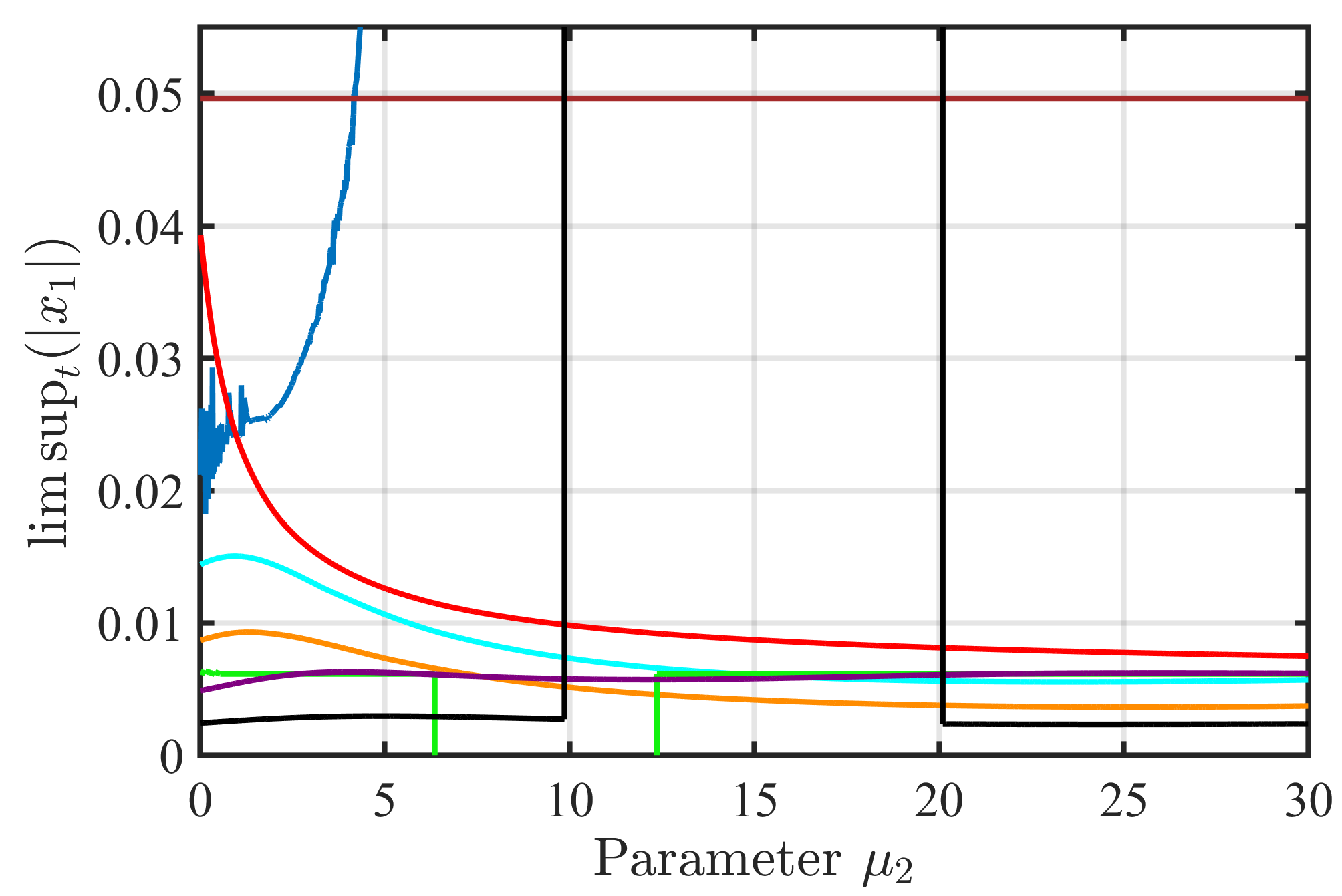}}
    \subfloat[\label{2d}][Accuracy over $\mu_{1}$ and $\mu_{2}$]{
        \includegraphics[scale=1.0]{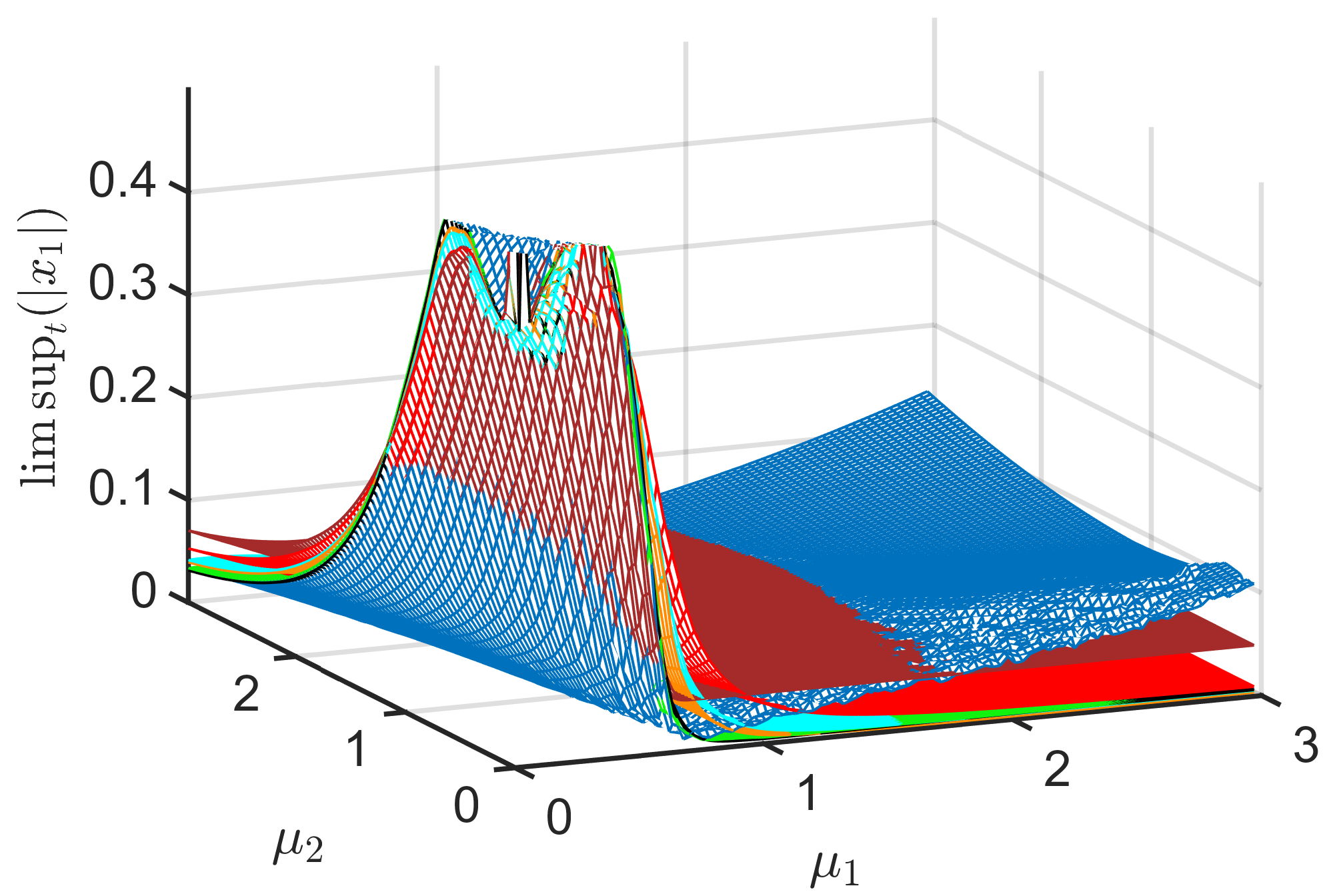}}
    \caption{Precision w.r.t. an increase of the controller's gains.}
    \label{fig:parametrs_effect}
\end{figure}


\subsection{Steady-State Accuracy when Varying the Controller Parameters}

Finally, simulations were performed regarding the steady-state accuracy when varying the controller parameters. In the simulation, all the parameters and the initial states are kept the same as in section \ref{sim_samp_time} except for the parameter to be changed. 

In the first simulation, controller gains are selected as ${{k}_{1}}=1.5\sqrt{\Lambda}$, ${{k}_{2}}=2.2\sqrt{\Lambda}$, ${{\mu }_{1}}=1$ and ${{\mu}_{2}}=1$. This parameter relation corresponds to the suggested parameter choice in Reference \citenum{koch2019discrete}. Figure \ref{fig:parametrs_effect}(a) depicts ${{e}_{f}}$ over varying $\Lambda$ from 0 to 60 in 1000 steps. The SSE of ${{\sum}_{\rm Explicit}}$ first drops quickly when $\Lambda$ is at a small value and then increases rapidly. Especially, the SSE even has a big jump when $\Lambda$ closes to 60. The result shows that the explicit method is very sensitive to overlarge controller gain. In contrast, the SSE of ${{\sum}_{\rm Implicit}}$ quickly decreases to a fixed value as $\Lambda$ increases. For the eigenvalue mapping-based method, the result illustrates that their precision won't deteriorate when increasing $\Lambda$ and can provide much higher precision than the Implicit Euler method. From the enlarged view of Figure \ref{fig:parametrs_effect}(a), it shows that the proposed ${{\sum }_{\rm Gudermannian}}$ achieves the highest precision when $\Lambda$ is large enough.   

Figure \ref{fig:parametrs_effect}(b) and Figure \ref{fig:parametrs_effect}(c) show the SSE over varying ${k}_{1}$ and ${k}_{2}$ from 1 to 60, respectively, in 1000 steps. The second controller parameter was fixed at 10 and the other settings were kept the same as in the first simulation. In Figure \ref{fig:parametrs_effect}(b), the SSE of ${{\sum }_{\rm Explicit}}$ first decreases and then increases rapidly to a large value which shows that the Explicit method is rather sensitive to gain ${k}_{1}$. The system ${{\sum }_{\rm Implicit}}$ is independent of gain ${k}_{1}$ and the SSE maintains at a fixed value. For the eigenvalue mapping-based method, the SSE falls from a large error quickly and then rises to the neighborhood of ${{\sum }_{\rm Implicit}}$. However, the SSE of ${{\sum }_{\rm Gudermannian}}$ finally settles at a much lower value than others, which shows its superiority. In Figure \ref{fig:parametrs_effect}(c), the SSE of ${{\sum }_{\rm Explicit}}$ still varies at a high level compared to others. The SSE of ${{\sum}_{\rm Implicit}}$ maintains at a high level. The SSE of eigenvalue mapping-based methods decreases to a much lower level than ${{\sum }_{\rm Implicit}}$. It can be observed that the proposed eigenvalue mapping methods, especially the Gudermannian method, achieve smaller SSE than the original ones. 

In the third group simulations, the SSE over $\mu_1$ and $\mu_2$ will be investigated. The results in Figure \ref{fig:parametrs_effect}(d) are very similar to \ref{fig:parametrs_effect}(b). It's worth noting that ${{\sum }_{\rm ReLU}}$ is unstable when ${\mu_{1}}$ increases to a specific interval. The SSE of the proposed methods decreases faster and settles at a lower level when increasing ${\mu_{1}}$. Furthermore, increasing ${\mu_{1}}$ drives the SSE of ${{\sum }_{\rm Gudermannian}}$ quickly converges to the lowest level which shows its strong robustness. From Figure \ref{fig:parametrs_effect}(e), it can be observed that the effect of ${\mu_{1}}$ on the SSE is similar to that of ${k_{2}}$. But ${{\sum }_{\rm ReLU}}$ and ${{\sum }_{\rm Gudermannian}}$ become unstable in a specific interval when increasing ${\mu_{2}}$. Finally, the common influence of $\mu_{1}$ and $\mu_{2}$ on the SSE is explored in Figure\ref{fig:parametrs_effect}(f). It can be seen that the surface of ${{\sum }_{\rm Explicit}}$ bends upward with the increasing of $\mu_{1}$ and $\mu_{2}$. The surfaces of the proposed methods are under the original methods after the gains increment to a certain value. The results show that the proposed methods are insensitive to overestimation of $\mu_{1}$ and $\mu_{2}$ and can achieve higher precision.

\section{CONCLUSIONS}
In this paper, a novel eigenvalue mapping-based discrete-time GSTA is presented. The existing EMFs are analyzed and four new EMFs are proposed to further enhance the control performance. All EMFs are extended to the complex domain which relaxes the restrictions on parameter selection. The obtained controllers are given by explicit recursions, guaranteeing a straightforward implementation in real applications. The proposed method is directly compared to previously published discrete-time STA in a series of simulation examples. The proposed method entirely avoids discretization chattering. Further, It can provide higher asymptotic accuracy w.r.t. the sampling time compared to other existing methods. It also demonstrates that the proposed method is insensitive to gain overestimation and the precision is improved when increasing the controller gains. The proposed method can provide higher steady-state accuracy than existing discretization methods. Future work will focus on the stability analysis that is not carried on in this article. We also want to investigate the impact of measurement noise on the control accuracy.

\nocite{*}
\bibliography{main}

\begin{thebibliography}{10}
\providecommand \doibase [0]{http://dx.doi.org/}%

\bibitem{weissenberger2022discretization}
Weissenberger F, Watermann L, Koch S, et al. Discretization of the Super-Twisting Algorithm Using Variational Integrators. In: IEEE. ; 2022\string: 4307--4312.

\bibitem{andritsch2023modified}
Andritsch B, Watermann L, Koch S, Reichhartinger M, Reger J, Horn M. Modified Implicit Discretization of the Super-Twisting Controller. {\it arXiv preprint arXiv:2303.15273} 2023.

\bibitem{yan2020euler}
Yan Y, Yu S, Yu X. Euler's discretization effect on a sliding-mode control system with supertwisting algorithm. {\it IEEE Transactions on Automatic Control} 2020\string; 66(6)\string: 2817--2824.

\bibitem{filippov2013differential}
Filippov AF. {\it Differential equations with discontinuous righthand sides: control systems}. 18.
\newblock Springer Science \& Business Media .
\newblock 2013.

\bibitem{levant2003higher}
Levant A. Higher-order sliding modes, differentiation and output-feedback control. {\it International journal of Control} 2003\string; 76(9-10)\string: 924--941.

\bibitem{cao2022adaptive}
Cao G, Yang J, Qiao L, Yang Z, Zhang W. Adaptive output feedback super twisting algorithm for trajectory tracking control of USVs with saturated constraints. {\it Ocean Engineering} 2022\string; 259\string: 111507.

\bibitem{zhang2022robust}
Zhang W, Du H, Chu Z. Robust discrete-time non-smooth consensus protocol for multi-agent systems via super-twisting algorithm. {\it Applied Mathematics and Computation} 2022\string; 413\string: 126636.

\bibitem{moreno2009linear}
Moreno JA. A linear framework for the robust stability analysis of a generalized super-twisting algorithm. In: IEEE. ; 2009\string: 1--6.

\bibitem{castillo2018super}
Castillo I, Fridman L, Moreno JA. Super-twisting algorithm in presence of time and state dependent perturbations. {\it International Journal of Control} 2018\string; 91(11)\string: 2535--2548.

\bibitem{borlaug2022generalized}
Borlaug ILG, Pettersen KY, Gravdahl JT. The generalized super-twisting algorithm with adaptive gains. {\it International Journal of Robust and Nonlinear Control} 2022\string; 32(13)\string: 7240--7270.

\bibitem{yan2015discretization}
Yan Y, Yu X, Sun C. Discretization behaviors of a super-twisting algorithm based sliding mode control system. In: IEEE. ; 2015\string: 1--5.

\bibitem{brogliato2019implicit}
Brogliato B, Polyakov A, Efimov D. The implicit discretization of the supertwisting sliding-mode control algorithm. {\it IEEE Transactions on Automatic Control} 2019\string; 65(8)\string: 3707--3713.

\bibitem{brogliato2021digital}
Brogliato B, Polyakov A. Digital implementation of sliding-mode control via the implicit method: A tutorial. {\it International Journal of Robust and Nonlinear Control} 2021\string; 31(9)\string: 3528--3586.

\bibitem{xiong2020discrete}
Xiong X, Liu Z, Kamal S, Jin S. Discrete-time super-twisting observer with implicit Euler method. {\it IEEE Transactions on Circuits and Systems II: Express Briefs} 2020\string; 68(4)\string: 1288--1292.

\bibitem{xiong2021discrete}
Xiong X, Chen G, Lou Y, Huang R, Kamal S. Discrete-Time Implementation of Super-Twisting Control With Semi-Implicit Euler Method. {\it IEEE Transactions on Circuits and Systems II: Express Briefs} 2021\string; 69(1)\string: 99--103.

\bibitem{koch2019discrete}
Koch S, Reichhartinger M. Discrete-time equivalents of the super-twisting algorithm. {\it Automatica} 2019\string; 107\string: 190--199.

\bibitem{koch2019discretization}
Koch S, Reichhartinger M, Horn M. On the discretization of the super-twisting algorithm. In: IEEE. ; 2019\string: 5989--5994.

\bibitem{eisenzopf2021adaptive}
Eisenzopf L, Watermann L, Koch S, Reichhartinger M, Reger J, Horn M. Adaptive Gain Super-Twisting-Algorithm: Design and Discretization. In: IEEE. ; 2021\string: 6415--6420.

\bibitem{reichhartinger2018arbitrary}
Reichhartinger M, Spurgeon S. An arbitrary-order differentiator design paradigm with adaptive gains. {\it International Journal of Control} 2018\string; 91(9)\string: 2028--2042.

\bibitem{huber2019lyapunov}
Huber O, Acary V, Brogliato B. Lyapunov stability analysis of the implicit discrete-time twisting control algorithm. {\it IEEE Transactions on Automatic Control} 2019\string; 65(6)\string: 2619--2626.

\bibitem{huber2016implicit}
Huber O, Acary V, Brogliato B, Plestan F. Implicit discrete-time twisting controller without numerical chattering: Analysis and experimental results. {\it Control Engineering Practice} 2016\string; 46\string: 129--141.

\bibitem{wetzlinger2019semi}
Wetzlinger M, Reichhartinger M, Horn M, Fridman L, Moreno JA. Semi-implicit discretization of the uniform robust exact differentiator. In: IEEE. ; 2019\string: 5995--6000.

\bibitem{ding2021station}
Ding N, Tang Y, Jiang Z, Bai Y, Liang S. Station-Keeping Control of Autonomous and Remotely-Operated Vehicles for Free Floating Manipulation. {\it Journal of Marine Science and Engineering} 2021\string; 9(11)\string: 1305.

\bibitem{mcclamroch2016stability}
McClamroch NH. Stability of Dynamical Systems-On the Role of Monotonic and Non-Monotonic Lyapunov Functions [Bookshelf]. {\it IEEE Control Systems Magazine} 2016\string; 36(1)\string: 77--78.

\bibitem{yu2003discretization}
Yu X, Chen G. Discretization behaviors of equivalent control based sliding-mode control systems. {\it IEEE Transactions on Automatic Control} 2003\string; 48(9)\string: 1641--1646.

\bibitem{prasun2023minimum}
Prasun P, Pandey S, Kamal S, Ghosh S, Xiong X. A Minimum Operator Based Discrete-Time Super-Twisting-Like Algorithm. {\it IEEE Transactions on Circuits and Systems II: Express Briefs} 2023.

\end{thebibliography}

\appendix
\section{Limitation of the matching-explicit EMF when the sampling time approximates zero
}

In order to facilitate typesetting, we abbreviate ${\lambda}_{n,k}(x_{1,k})$ to ${\lambda}_{n}$ with $i=1,2$. The rest of the APPENDIX also follows this abbreviation. 

\begin{eqnarray}
  && \underset{h\to 0}{\mathop{\lim }}\,\frac{1}{h}\left( {{q}_{1}}+{{q}_{2}}-2 \right)=\underset{h\to 0}{\mathop{\lim }}\,\frac{1}{h}\left( {{e}^{\frac{1}{2}h{{\lambda }_{1}}}}\left( 1+\tfrac{1}{2}h{{\lambda }_{1}} \right)+{{e}^{\frac{1}{2}h{{\lambda }_{2}}}}\left( 1+\tfrac{1}{2}h{{\lambda }_{2}} \right)-2 \right) \nonumber\\ 
 && =\underset{h\to 0}{\mathop{\lim }}\,\frac{{{e}^{\frac{1}{2}h{{\lambda }_{1}}}}\left( 1+\tfrac{1}{2}h{{\lambda }_{1}} \right)+{{e}^{\frac{1}{2}h{{\lambda }_{2}}}}\left( 1+\tfrac{1}{2}h{{\lambda }_{2}} \right)-2}{h} \nonumber\\ 
 && =\underset{h\to 0}{\mathop{\lim }}\,\tfrac{1}{2}{{\lambda }_{1}}{{e}^{\tfrac{1}{2}h{{\lambda }_{1}}}}\left( 1+\tfrac{1}{2}h{{\lambda }_{1}} \right)+\tfrac{1}{2}{{\lambda }_{1}}{{e}^{\tfrac{1}{2}h{{\lambda }_{1}}}}+\tfrac{1}{2}{{\lambda }_{2}}{{e}^{\tfrac{1}{2}h{{\lambda }_{2}}}}\left( 1+\tfrac{1}{2}h{{\lambda }_{2}} \right)+\tfrac{1}{2}{{\lambda }_{2}}{{e}^{\tfrac{1}{2}h{{\lambda }_{2}}}} \nonumber\\
 && =\tfrac{1}{2}{{\lambda }_{1}}+\tfrac{1}{2}{{\lambda }_{1}}+\tfrac{1}{2}{{\lambda }_{2}}+\tfrac{1}{2}{{\lambda }_{2}} \nonumber\\
 && ={{\lambda }_{1}}+{{\lambda }_{2}}
\end{eqnarray}

\begin{eqnarray}
  && \underset{h\to 0}{\mathop{\lim }}\,\frac{1}{{{h}^{2}}}\left( {{q}_{1}}+{{q}_{2}}-{{q}_{1}}{{q}_{2}}-1 \right)\nonumber\\
  && =\underset{h\to 0}{\mathop{\lim }}\,\frac{{{e}^{\frac{1}{2}h{{\lambda }_{1}}}}\left( 1+\tfrac{1}{2}h{{\lambda }_{1}} \right)+{{e}^{\frac{1}{2}h{{\lambda }_{2}}}}\left( 1+\tfrac{1}{2}h{{\lambda }_{2}} \right)-{{e}^{\frac{1}{2}h\left( {{\lambda }_{1}}+{{\lambda }_{2}} \right)}}\left( 1+\tfrac{1}{2}h{{\lambda }_{1}} \right)\left( 1+\tfrac{1}{2}h{{\lambda }_{2}} \right)-1}{{{h}^{2}}} \nonumber\\
  && =\underset{h\to 0}{\mathop{\lim }}\,\frac{\begin{split}
  & \tfrac{1}{2}{{\lambda }_{1}}{{e}^{\frac{1}{2}h{{\lambda }_{1}}}}\left( 1+\tfrac{1}{2}h{{\lambda }_{1}} \right)+\tfrac{1}{2}{{\lambda }_{1}}{{e}^{\frac{1}{2}h{{\lambda }_{1}}}}+\tfrac{1}{2}{{\lambda }_{2}}{{e}^{\frac{1}{2}h{{\lambda }_{2}}}}\left( 1+\tfrac{1}{2}h{{\lambda }_{2}} \right)+\tfrac{1}{2}{{\lambda }_{2}}{{e}^{\frac{1}{2}h{{\lambda }_{2}}}} \\ 
  & -{{e}^{\frac{1}{2}h\left( {{\lambda }_{1}}+{{\lambda }_{2}} \right)}}\left[ \tfrac{1}{2}\left( {{\lambda }_{1}}+{{\lambda }_{2}} \right)\left( 1+\tfrac{1}{2}h{{\lambda }_{1}} \right)\left( 1+\tfrac{1}{2}h{{\lambda }_{2}} \right)+\tfrac{1}{2}{{\lambda }_{1}}\left( 1+\tfrac{1}{2}h{{\lambda }_{2}} \right)+\tfrac{1}{2}{{\lambda }_{2}}\left( 1+\tfrac{1}{2}h{{\lambda }_{1}} \right) \right] 
  \end{split}}{2h} \nonumber\\ 
  && =\underset{h\to 0}{\mathop{\lim }}\,\frac{\begin{split}
  & \tfrac{1}{4}\lambda _{1}^{2}{{e}^{\frac{1}{2}h{{\lambda }_{1}}}}\left( 1+\tfrac{1}{2}h{{\lambda }_{1}} \right)+\tfrac{1}{4}\lambda _{1}^{2}{{e}^{\frac{1}{2}h{{\lambda }_{1}}}}+\tfrac{1}{4}\lambda _{2}^{2}{{e}^{\frac{1}{2}h{{\lambda }_{2}}}}\left( 1+\tfrac{1}{2}h{{\lambda }_{2}} \right)+\tfrac{1}{4}\lambda _{2}^{2}{{e}^{\frac{1}{2}h{{\lambda }_{2}}}}+\tfrac{1}{4}\lambda _{1}^{2}{{e}^{\frac{1}{2}h{{\lambda }_{1}}}}+\tfrac{1}{4}\lambda _{2}^{2}{{e}^{\frac{1}{2}h{{\lambda }_{2}}}} \\ 
 & -\tfrac{1}{2}\left( {{\lambda }_{1}}+{{\lambda }_{2}} \right){{e}^{\frac{1}{2}h\left( {{\lambda }_{1}}+{{\lambda }_{2}} \right)}}\left[ \tfrac{1}{2}\left( {{\lambda }_{1}}+{{\lambda }_{2}} \right)\left( 1+\tfrac{1}{2}h{{\lambda }_{1}} \right)\left( 1+\tfrac{1}{2}h{{\lambda }_{2}} \right)+\tfrac{1}{2}{{\lambda }_{1}}\left( 1+\tfrac{1}{2}h{{\lambda }_{2}} \right)+\tfrac{1}{2}{{\lambda }_{2}}\left( 1+\tfrac{1}{2}h{{\lambda }_{1}} \right) \right] \\ 
 & -{{e}^{\frac{1}{2}h\left( {{\lambda }_{1}}+{{\lambda }_{2}} \right)}}\left[ \tfrac{1}{4}{{\lambda }_{1}}\left( {{\lambda }_{1}}+{{\lambda }_{2}} \right)\left( 1+\tfrac{1}{2}h{{\lambda }_{2}} \right)+\tfrac{1}{4}{{\lambda }_{2}}\left( {{\lambda }_{1}}+{{\lambda }_{2}} \right)\left( 1+\tfrac{1}{2}h{{\lambda }_{1}} \right)+\tfrac{1}{4}{{\lambda }_{1}}{{\lambda }_{2}}+\tfrac{1}{4}{{\lambda }_{1}}{{\lambda }_{2}} \right] \\ 
 \end{split}}{2}
 \nonumber\\
 && =-{{\lambda }_{1}}{{\lambda }_{2}}
\end{eqnarray}


\begin{eqnarray}
    && \underset{h\to 0}{\mathop{\lim }}\,\frac{1}{h}\left( {{q}_{1}}(h,{{x}_{1,k}})+{{q}_{2}}(h,{{x}_{1,k}})-2 \right) \nonumber\\ 
    && =\underset{h\to 0}{\mathop{\lim }}\,\frac{{{e}^{\frac{1}{2}h\left( {{\lambda }_{r}}+{{\lambda }_{i}}i \right)}}\left[ 1+\tfrac{1}{2}h\left( {{\lambda }_{r}}+{{\lambda }_{i}}i \right) \right]+{{e}^{\frac{1}{2}h\left( {{\lambda }_{r}}-{{\lambda }_{i}}i \right)}}\left[ 1+\tfrac{1}{2}h\left( {{\lambda }_{r}}-{{\lambda }_{i}}i \right) \right]-2}{h} \nonumber\\ 
    && =\underset{h\to 0}{\mathop{\lim }}\,\frac{{{e}^{\frac{1}{2}h{{\lambda }_{r}}}}\left[ 2\cos \left( \tfrac{1}{2}h{{\lambda }_{i}} \right)\left( 1+\tfrac{1}{2}h{{\lambda }_{r}} \right)-h{{\lambda }_{i}}\sin \left( \tfrac{1}{2}h{{\lambda }_{i}} \right) \right]-2}{h} \nonumber\\
    && ={{e}^{\frac{1}{2}h{{\lambda }_{r}}}}\left[ {{\lambda }_{r}}\cos \tfrac{1}{2}h{{\lambda }_{i}}\left( 2+\tfrac{1}{2}h{{\lambda }_{r}} \right)-\tfrac{1}{2}h{{\lambda }_{r}}{{\lambda }_{i}}\sin \tfrac{1}{2}h{{\lambda }_{i}}-{{\lambda }_{i}}\sin \left( \tfrac{1}{2}h{{\lambda }_{i}} \right)\left( 2+\tfrac{1}{2}h{{\lambda }_{r}} \right)-\tfrac{1}{2}h\lambda _{i}^{2}\cos \left( \tfrac{1}{2}h{{\lambda }_{i}} \right) \right] \nonumber\\
    && =2{{\lambda }_{r}}
\end{eqnarray}

\begin{eqnarray}
    && \underset{h\to 0}{\mathop{\lim }}\,\frac{1}{{{h}^{2}}}\left( {{q}_{1}}+{{q}_{2}}-{{q}_{1}}{{q}_{2}}-1 \right) \nonumber\\ 
    && =\underset{h\to 0}{\mathop{\lim }}\,\frac{{{e}^{\frac{1}{2}h\left( {{\lambda }_{r}}+{{\lambda }_{i}}i \right)}}\left[ 1+\tfrac{1}{2}h\left( {{\lambda }_{r}}+{{\lambda }_{i}}i \right) \right]+{{e}^{\frac{1}{2}h\left( {{\lambda }_{r}}-{{\lambda }_{i}}i \right)}}\left[ 1+\tfrac{1}{2}h\left( {{\lambda }_{r}}-{{\lambda }_{i}}i \right) \right]-{{e}^{h{{\lambda }_{r}}}}\left[ 1+\tfrac{1}{2}h\left( {{\lambda }_{r}}+{{\lambda }_{i}}i \right) \right]\left[ 1+\tfrac{1}{2}h\left( {{\lambda }_{r}}-{{\lambda }_{i}}i \right) \right]-1}{{{h}^{2}}} \nonumber\\ 
    && =\underset{h\to 0}{\mathop{\lim }}\,\frac{{{e}^{\frac{1}{2}h{{\lambda }_{r}}}}\left[ 2\cos \left( \tfrac{1}{2}h{{\lambda }_{i}} \right)\left( 1+\tfrac{1}{2}h{{\lambda }_{r}} \right)-\sin \left( \tfrac{1}{2}h{{\lambda }_{i}} \right)h{{\lambda }_{i}} \right]-{{e}^{h{{\lambda }_{r}}}}\left[ {{\left( 1+\tfrac{1}{2}h{{\lambda }_{r}} \right)}^{2}}+\tfrac{1}{4}{{h}^{2}}\lambda _{i}^{2} \right]-1}{{{h}^{2}}} \nonumber\\
    &&=\underset{h\to 0}{\mathop{\lim }}\,\frac{
    \begin{split}
     &{{e}^{\frac{1}{2}h{{\lambda }_{r}}}}\left[ \left( 2{{\lambda }_{r}}+\tfrac{1}{2}h\lambda _{r}^{2}-\tfrac{1}{2}h\lambda _{i}^{2} \right)\cos \left( \tfrac{1}{2}h{{\lambda }_{i}} \right)-\left( 2{{\lambda }_{i}}+h{{\lambda }_{r}}{{\lambda }_{i}} \right)\sin \left( \tfrac{1}{2}h{{\lambda }_{i}} \right) \right] \\ 
     &-{{e}^{h{{\lambda }_{r}}}}\left[ {{\lambda }_{r}}{{\left( 1+\tfrac{1}{2}h{{\lambda }_{r}} \right)}^{2}}+\tfrac{1}{4}{{h}^{2}}\lambda _{i}^{2}{{\lambda }_{r}}+{{\lambda }_{r}}\left( 1+\tfrac{1}{2}h{{\lambda }_{r}} \right)+\tfrac{1}{2}h\lambda _{i}^{2} \right]
    \end{split}}{2h} \nonumber\\
    &&=\underset{h\to 0}{\mathop{\lim }}\,\frac{\begin{split}
      & {{e}^{\frac{1}{2}h{{\lambda }_{r}}}}\left[ \left( \tfrac{3}{2}\lambda _{r}^{2}-\tfrac{3}{2}\lambda _{i}^{2}+\tfrac{1}{4}h\lambda _{r}^{3}-\tfrac{3}{4}h{{\lambda }_{r}}\lambda _{i}^{2} \right)\cos \left( \tfrac{1}{2}h{{\lambda }_{i}} \right)-\left( 3{{\lambda }_{r}}{{\lambda }_{i}}+\tfrac{3}{4}h\lambda _{r}^{2}{{\lambda }_{i}}-\tfrac{1}{4}h\lambda _{i}^{3} \right)\sin \left( \tfrac{1}{2}h{{\lambda }_{i}} \right) \right] \\ 
     & -{{e}^{h{{\lambda }_{r}}}}\left[ \lambda _{r}^{2}{{\left( 1+\tfrac{1}{2}h{{\lambda }_{r}} \right)}^{2}}+\tfrac{1}{4}{{h}^{2}}\lambda _{i}^{2}\lambda _{r}^{2}+2\lambda _{r}^{2}\left( 1+\tfrac{1}{2}h{{\lambda }_{r}} \right)+h\lambda _{i}^{2}{{\lambda }_{r}}+\tfrac{1}{2}\lambda _{r}^{2}+\tfrac{1}{2}\lambda _{i}^{2} \right] \\ 
    \end{split}}{2} \nonumber\\
    &&=\underset{h\to 0}{\mathop{\lim }}\,\frac{\tfrac{3}{2}\lambda _{r}^{2}-\tfrac{3}{2}\lambda _{i}^{2}+\tfrac{1}{4}h\lambda _{r}^{3}-\tfrac{3}{4}h{{\lambda }_{r}}\lambda _{i}^{2}-\left( \lambda _{r}^{2}+2\lambda _{r}^{2}+\tfrac{1}{2}\lambda _{r}^{2}+\tfrac{1}{2}\lambda _{i}^{2} \right)}{2}\nonumber\\
    && =-\lambda _{r}^{2}-\lambda _{i}^{2}
\end{eqnarray}

\section{Limitation of the tanh EMF when the sampling time approximates zero}

\begin{eqnarray}
  && \underset{h\to 0}{\mathop{\lim }}\,\frac{1}{h}\left( {{q}_{1}}+{{q}_{2}}-2 \right)=\underset{h\to 0}{\mathop{\lim }}\,\frac{1}{h}\left( \frac{\text{2}{{e}^{h{{\lambda }_{1}}}}}{{{e}^{h{{\lambda }_{1}}}}\text{+}{{e}^{-h{{\lambda }_{1}}}}}+\frac{\text{2}{{e}^{h{{\lambda }_{2}}}}}{{{e}^{h{{\lambda }_{2}}}}\text{+}{{e}^{-h{{\lambda }_{2}}}}}-2 \right) \nonumber\\ 
 && =\underset{h\to 0}{\mathop{\lim }}\,\frac{2{{e}^{h{{\lambda }_{1}}+h{{\lambda }_{2}}}}-\text{2}{{e}^{-h{{\lambda }_{1}}-h{{\lambda }_{2}}}}}{h\left( {{e}^{h{{\lambda }_{1}}+h{{\lambda }_{2}}}}\text{+}{{e}^{-h{{\lambda }_{1}}-h{{\lambda }_{2}}}}\text{+}{{e}^{h{{\lambda }_{1}}-h{{\lambda }_{2}}}}+{{e}^{-h{{\lambda }_{1}}+h{{\lambda }_{2}}}} \right)} \nonumber\\ 
 && =\underset{h\to 0}{\mathop{\lim }}\,\frac{2({{\lambda }_{1}}+{{\lambda }_{2}})\left( {{e}^{h{{\lambda }_{1}}+h{{\lambda }_{2}}}}\text{+}{{e}^{-h{{\lambda }_{1}}-h{{\lambda }_{2}}}} \right)}{{{e}^{h{{\lambda }_{1}}+h{{\lambda }_{2}}}}\text{+}{{e}^{h{{\lambda }_{1}}-h{{\lambda }_{2}}}}+{{e}^{-h{{\lambda }_{1}}+h{{\lambda }_{2}}}}\text{+}{{e}^{-h{{\lambda }_{1}}-h{{\lambda }_{2}}}}+h\left[ \left( {{\lambda }_{1}}+{{\lambda }_{2}} \right)\left( {{e}^{h{{\lambda }_{1}}+h{{\lambda }_{2}}}}-{{e}^{-h{{\lambda }_{1}}-h{{\lambda }_{2}}}} \right)\text{+}\left( {{\lambda }_{1}}-{{\lambda }_{2}} \right)\left( {{e}^{h{{\lambda }_{1}}-h{{\lambda }_{2}}}}-{{e}^{-h{{\lambda }_{1}}+h{{\lambda }_{2}}}} \right) \right]} \nonumber\\ 
 && ={{\lambda }_{1}}+{{\lambda }_{2}}
\end{eqnarray}

\begin{eqnarray}
  && \underset{h\to 0}{\mathop{\lim }}\,\frac{1}{{{h}^{2}}}\left( {{q}_{1}}+{{q}_{2}}-{{q}_{1}}{{q}_{2}}-1 \right)\text{=}\underset{h\to 0}{\mathop{\lim }}\,\frac{1}{{{h}^{2}}}\left( \frac{\text{2}{{e}^{h{{\lambda }_{1}}}}}{{{e}^{h{{\lambda }_{1}}}}\text{+}{{e}^{-h{{\lambda }_{1}}}}}+\frac{\text{2}{{e}^{h{{\lambda }_{2}}}}}{{{e}^{h{{\lambda }_{2}}}}\text{+}{{e}^{-h{{\lambda }_{2}}}}}-\frac{\text{2}{{e}^{h{{\lambda }_{1}}}}}{{{e}^{h{{\lambda }_{1}}}}\text{+}{{e}^{-h{{\lambda }_{1}}}}}\frac{\text{2}{{e}^{h{{\lambda }_{2}}}}}{{{e}^{h{{\lambda }_{2}}}}\text{+}{{e}^{-h{{\lambda }_{2}}}}}-1 \right) \nonumber\\ 
 && =\underset{h\to 0}{\mathop{\lim }}\,\frac{{{e}^{h{{\lambda }_{1}}-h{{\lambda }_{2}}}}\text{+}{{e}^{-h{{\lambda }_{1}}+h{{\lambda }_{2}}}}-{{e}^{h{{\lambda }_{1}}+h{{\lambda }_{2}}}}-{{e}^{-h{{\lambda }_{1}}-h{{\lambda }_{2}}}}}{{{h}^{2}}\left( {{e}^{h{{\lambda }_{1}}+h{{\lambda }_{2}}}}\text{+}{{e}^{h{{\lambda }_{1}}-h{{\lambda }_{2}}}}+{{e}^{-h{{\lambda }_{1}}+h{{\lambda }_{2}}}}\text{+}{{e}^{-h{{\lambda }_{1}}-h{{\lambda }_{2}}}} \right)} \nonumber\\ 
 && =\underset{h\to 0}{\mathop{\lim }}\,\frac{({{\lambda }_{1}}-{{\lambda }_{2}}){{e}^{h{{\lambda }_{1}}-h{{\lambda }_{2}}}}\text{+(}-{{\lambda }_{1}}+{{\lambda }_{2}}\text{)}{{e}^{-h{{\lambda }_{1}}+h{{\lambda }_{2}}}}-({{\lambda }_{1}}+{{\lambda }_{2}}){{e}^{h{{\lambda }_{1}}+h{{\lambda }_{2}}}}+({{\lambda }_{1}}+{{\lambda }_{2}}){{e}^{-h{{\lambda }_{1}}-h{{\lambda }_{2}}}}}{2h\left( {{e}^{h{{\lambda }_{1}}+h{{\lambda }_{2}}}}\text{+}{{e}^{h{{\lambda }_{1}}-h{{\lambda }_{2}}}}+{{e}^{-h{{\lambda }_{1}}+h{{\lambda }_{2}}}}\text{+}{{e}^{-h{{\lambda }_{1}}-h{{\lambda }_{2}}}} \right)+{{h}^{2}}\left[ \left( {{\lambda }_{1}}+{{\lambda }_{2}} \right)\left( {{e}^{h{{\lambda }_{1}}+h{{\lambda }_{2}}}}-{{e}^{-h{{\lambda }_{1}}-h{{\lambda }_{2}}}} \right)\text{+}\left( {{\lambda }_{1}}-{{\lambda }_{2}} \right)\left( {{e}^{h{{\lambda }_{1}}-h{{\lambda }_{2}}}}-{{e}^{-h{{\lambda }_{1}}+h{{\lambda }_{2}}}} \right) \right]} \nonumber\\ 
 && =\underset{h\to 0}{\mathop{\lim }}\,\frac{({{\lambda }_{1}}-{{\lambda }_{2}}){{e}^{h{{\lambda }_{1}}-h{{\lambda }_{2}}}}\text{+(}-{{\lambda }_{1}}+{{\lambda }_{2}}\text{)}{{e}^{-h{{\lambda }_{1}}+h{{\lambda }_{2}}}}-({{\lambda }_{1}}+{{\lambda }_{2}}){{e}^{h{{\lambda }_{1}}+h{{\lambda }_{2}}}}+({{\lambda }_{1}}+{{\lambda }_{2}}){{e}^{-h{{\lambda }_{1}}-h{{\lambda }_{2}}}}}{2\left( {{e}^{h{{\lambda }_{1}}+h{{\lambda }_{2}}}}\text{+}{{e}^{h{{\lambda }_{1}}-h{{\lambda }_{2}}}}+{{e}^{-h{{\lambda }_{1}}+h{{\lambda }_{2}}}}\text{+}{{e}^{-h{{\lambda }_{1}}-h{{\lambda }_{2}}}} \right)+o(h)} \nonumber\\ 
 && =\underset{h\to 0}{\mathop{\lim }}\,\frac{{{\left( {{\lambda }_{1}}-{{\lambda }_{2}} \right)}^{2}}\text{+}{{\left( {{\lambda }_{1}}-{{\lambda }_{2}} \right)}^{2}}-{{\left( {{\lambda }_{1}}+{{\lambda }_{2}} \right)}^{2}}-{{\left( {{\lambda }_{1}}+{{\lambda }_{2}} \right)}^{2}}}{8} \nonumber\\ 
 && =-{{\lambda }_{1}}{{\lambda }_{2}}
\end{eqnarray}

\begin{eqnarray}
  && \underset{h\to 0}{\mathop{\lim }}\,\frac{1}{h}\left( {{q}_{1}}+{{q}_{2}}-2 \right)=\underset{h\to 0}{\mathop{\lim }}\,\frac{\text{2}{{e}^{h\left( {{\lambda }_{r}}+{{\lambda }_{i}}i \right)}}}{{{e}^{h\left( {{\lambda }_{r}}+{{\lambda }_{i}}i \right)}}\text{+}{{e}^{-h\left( {{\lambda }_{r}}+{{\lambda }_{i}}i \right)}}}+\frac{\text{2}{{e}^{h\left( {{\lambda }_{r}}-{{\lambda }_{i}}i \right)}}}{{{e}^{h\left( {{\lambda }_{r}}-{{\lambda }_{i}}i \right)}}\text{+}{{e}^{-h\left( {{\lambda }_{r}}-{{\lambda }_{i}}i \right)}}}-2 \nonumber\\ 
 && =\underset{h\to 0}{\mathop{\lim }}\,\frac{2{{e}^{2h{{\lambda }_{r}}}}-\text{2}{{e}^{-2h{{\lambda }_{r}}}}}{h\left[ {{e}^{2h{{\lambda }_{r}}}}\text{+}{{e}^{-2h{{\lambda }_{r}}}}\text{+}2\cos (2h{{\lambda }_{i}}) \right]} \nonumber\\ 
 && =\underset{h\to 0}{\mathop{\lim }}\,\frac{4{{\lambda }_{r}}{{e}^{2h{{\lambda }_{r}}}}+4{{\lambda }_{r}}{{e}^{-2h{{\lambda }_{r}}}}}{{{e}^{2h{{\lambda }_{r}}}}\text{+}{{e}^{-2h{{\lambda }_{r}}}}\text{+}2\cos (2h{{\lambda }_{i}})+h\left[ 2{{\lambda }_{r}}{{e}^{2h{{\lambda }_{r}}}}-2{{\lambda }_{r}}{{e}^{-2h{{\lambda }_{r}}}}-4{{\lambda }_{i}}\sin (2h{{\lambda }_{i}}) \right]} \nonumber\\ 
 && =2{{\lambda }_{r}} 
\end{eqnarray}

\begin{eqnarray}
  && \underset{h\to 0}{\mathop{\lim }}\,\frac{1}{{{h}^{2}}}\left( {{q}_{1}}+{{q}_{2}}-{{q}_{1}}{{q}_{2}}-1 \right) \nonumber\\ 
 && =\underset{h\to 0}{\mathop{\lim }}\,\frac{\text{2}{{e}^{h\left( {{\lambda }_{r}}+{{\lambda }_{i}}i \right)}}}{{{e}^{h\left( {{\lambda }_{r}}+{{\lambda }_{i}}i \right)}}\text{+}{{e}^{-h\left( {{\lambda }_{r}}+{{\lambda }_{i}}i \right)}}}+\frac{\text{2}{{e}^{h\left( {{\lambda }_{r}}-{{\lambda }_{i}}i \right)}}}{{{e}^{h\left( {{\lambda }_{r}}-{{\lambda }_{i}}i \right)}}\text{+}{{e}^{-h\left( {{\lambda }_{r}}-{{\lambda }_{i}}i \right)}}}-\frac{\text{2}{{e}^{h\left( {{\lambda }_{r}}+{{\lambda }_{i}}i \right)}}}{{{e}^{h\left( {{\lambda }_{r}}+{{\lambda }_{i}}i \right)}}\text{+}{{e}^{-h\left( {{\lambda }_{r}}+{{\lambda }_{i}}i \right)}}}\frac{\text{2}{{e}^{h\left( {{\lambda }_{r}}-{{\lambda }_{i}}i \right)}}}{{{e}^{h\left( {{\lambda }_{r}}-{{\lambda }_{i}}i \right)}}\text{+}{{e}^{-h\left( {{\lambda }_{r}}-{{\lambda }_{i}}i \right)}}}-1 \nonumber\\ 
 && =\underset{h\to 0}{\mathop{\lim }}\,\frac{2\cos (2h{{\lambda }_{i}})-{{e}^{2h{{\lambda }_{r}}}}-{{e}^{-2h{{\lambda }_{r}}}}}{{{h}^{2}}\left[ {{e}^{2h{{\lambda }_{r}}}}\text{+}{{e}^{-2h{{\lambda }_{r}}}}\text{+}2\cos (2h{{\lambda }_{i}}) \right]} \nonumber\\ 
 && =\underset{h\to 0}{\mathop{\lim }}\,\frac{-4{{\lambda }_{i}}\sin (2h{{\lambda }_{i}})-2{{\lambda }_{r}}{{e}^{2h{{\lambda }_{r}}}}+2{{\lambda }_{r}}{{e}^{-2h{{\lambda }_{r}}}}}{2h\left[ {{e}^{2h{{\lambda }_{r}}}}\text{+}{{e}^{-2h{{\lambda }_{r}}}}\text{+}2\cos (2h{{\lambda }_{i}}) \right]+{{h}^{2}}\left[ 2{{\lambda }_{r}}{{e}^{2h{{\lambda }_{r}}}}-2{{\lambda }_{r}}{{e}^{-2h{{\lambda }_{r}}}}-4{{\lambda }_{i}}\sin (2h{{\lambda }_{i}}) \right]} \nonumber\\ 
 && =\underset{h\to 0}{\mathop{\lim }}\,\frac{-8\lambda _{i}^{2}\cos (2h{{\lambda }_{i}})-4\lambda _{r}^{2}{{e}^{2h{{\lambda }_{r}}}}-4\lambda _{r}^{2}{{e}^{-2h{{\lambda }_{r}}}}}{2\left[ {{e}^{2h{{\lambda }_{r}}}}\text{+}{{e}^{-2h{{\lambda }_{r}}}}\text{+}2\cos (2h{{\lambda }_{i}}) \right]+4h\left[ 2{{\lambda }_{r}}\left( {{e}^{2h{{\lambda }_{r}}}}-{{e}^{-2h{{\lambda }_{r}}}} \right)-4{{\lambda }_{i}}\sin (2h{{\lambda }_{i}}) \right]+o\left( h \right)} \nonumber\\ 
 && =-\lambda _{r}^{2}-\lambda _{i}^{2} 
\end{eqnarray}

\section{Limitation of the Gudermannian EMF when the sampling time approximates zero}

\begin{eqnarray}
  && \underset{h\to 0}{\mathop{\lim }}\,\frac{1}{h}\left( {{q}_{1}}+{{q}_{2}}-2 \right) \nonumber\\ 
 && =\underset{h\to 0}{\mathop{\lim }}\,\frac{\arctan \left( \sinh \left( h{{\lambda }_{r}}+ih{{\lambda }_{i}} \right) \right)+\arctan \left( \sinh \left( h{{\lambda }_{r}}-ih{{\lambda }_{i}} \right) \right)}{h} \nonumber\\ 
 && =\underset{h\to 0}{\mathop{\lim }}\,\frac{2\arctan \left( \frac{\sinh \left( h{{\lambda }_{r}} \right)}{\cos \left( h{{\lambda }_{i}} \right)} \right)}{h} \nonumber\\
 && =\underset{h\to 0}{\mathop{\lim }}\,2\frac{1}{1+{{\left( \frac{\sinh \left( h{{\lambda }_{r}} \right)}{\cos \left( h{{\lambda }_{i}} \right)} \right)}^{2}}}\frac{{{\lambda }_{r}}\cosh \left( h{{\lambda }_{r}} \right)\cos \left( h{{\lambda }_{i}} \right)+{{\lambda }_{i}}\sin \left( h{{\lambda }_{i}} \right)\sinh \left( h{{\lambda }_{r}} \right)}{{{\cos }^{2}}\left( h{{\lambda }_{i}} \right)} \nonumber\\
 && =2{{\lambda }_{r}} 
\end{eqnarray}

\begin{eqnarray}
  && \underset{h\to 0}{\mathop{\lim }}\,\frac{1}{{{h}^{2}}}\left( {{q}_{1}}+{{q}_{2}}-{{q}_{1}}{{q}_{2}}-1 \right) \nonumber\\ 
 && =\underset{h\to 0}{\mathop{\lim }}\,-\frac{\arctan \left( \sinh \left( h{{\lambda }_{r}}+ih{{\lambda }_{i}} \right) \right)\arctan \left( \sinh \left( h{{\lambda }_{r}}-ih{{\lambda }_{i}} \right) \right)}{{{h}^{2}}} \nonumber\\
 && =\underset{h\to 0}{\mathop{\lim }}\,-\frac{\left( \arctan \left( \frac{\sinh \left( h{{\lambda }_{r}} \right)}{\cos \left( h{{\lambda }_{i}} \right)} \right)+i\artanh \left( \frac{\sin \left( h{{\lambda }_{i}} \right)}{\cosh \left( h{{\lambda }_{r}} \right)} \right) \right)\left( \arctan \left( \frac{\sinh \left( h{{\lambda }_{r}} \right)}{\cos \left( h{{\lambda }_{i}} \right)} \right)-i\artanh \left( \frac{\sin \left( h{{\lambda }_{i}} \right)}{\cosh \left( h{{\lambda }_{r}} \right)} \right) \right)}{{{h}^{2}}} \nonumber\\
 && =\underset{h\to 0}{\mathop{\lim }}\,-\frac{{{\arctan }^{2}}\left( \frac{\sinh \left( h{{\lambda }_{r}} \right)}{\cos \left( h{{\lambda }_{i}} \right)} \right)+ {{\artanh}^{2}}\left( \frac{\sin \left( h{{\lambda }_{i}} \right)}{\cosh \left( h{{\lambda }_{r}} \right)} \right)}{{{h}^{2}}} \nonumber\\
 && =\underset{h\to 0}{\mathop{\lim }}\,-\frac{\arctan \left( \frac{\sinh \left( h{{\lambda }_{r}} \right)}{\cos \left( h{{\lambda }_{i}} \right)} \right)\frac{{{\lambda }_{r}}\cosh \left( h{{\lambda }_{r}} \right)\cos \left( h{{\lambda }_{i}} \right)+{{\lambda }_{i}}\sinh \left( h{{\lambda }_{r}} \right)\sin \left( h{{\lambda }_{i}} \right)}{{{\sinh }^{2}}\left( h{{\lambda }_{r}} \right)+{{\cos }^{2}}\left( h{{\lambda }_{i}} \right)}+\artanh \left( \frac{\sin \left( h{{\lambda }_{i}} \right)}{\cosh \left( h{{\lambda }_{r}} \right)} \right)\frac{{{\lambda }_{i}}\cos \left( h{{\lambda }_{i}} \right)\cosh \left( h{{\lambda }_{r}} \right)-{{\lambda }_{r}}\sin \left( h{{\lambda }_{i}} \right)\sinh \left( h{{\lambda }_{r}} \right)}{{{\cosh }^{2}}\left( h{{\lambda }_{r}} \right)-{{\sin }^{2}}\left( h{{\lambda }_{i}} \right)}}{h} \nonumber\\
 && =\underset{h\to 0}{\mathop{\lim }}\,-{{\left( \frac{{{\lambda }_{r}}\cosh \left( h{{\lambda }_{r}} \right)\cos \left( h{{\lambda }_{i}} \right)+{{\lambda }_{i}}\sinh \left( h{{\lambda }_{r}} \right)\sin \left( h{{\lambda }_{i}} \right)}{{{\sinh }^{2}}\left( h{{\lambda }_{r}} \right)+{{\cos }^{2}}\left( h{{\lambda }_{i}} \right)} \right)}^{2}}-\arctan \left( \frac{\sinh \left( h{{\lambda }_{r}} \right)}{\cos \left( h{{\lambda }_{i}} \right)} \right)\left( \cdots  \right) \nonumber\\
 && -{{\left( \frac{{{\lambda }_{i}}\cos \left( h{{\lambda }_{i}} \right)\cosh \left( h{{\lambda }_{r}} \right)-{{\lambda }_{r}}\sin \left( h{{\lambda }_{i}} \right)\sinh \left( h{{\lambda }_{r}} \right)}{{{\cosh }^{2}}\left( h{{\lambda }_{r}} \right)-{{\sin }^{2}}\left( h{{\lambda }_{i}} \right)} \right)}^{2}}-\artanh\left( \frac{\sin \left( h{{\lambda }_{i}} \right)}{\cosh \left( h{{\lambda }_{r}} \right)} \right)\left( \cdots  \right) \nonumber\\
 && =-\lambda _{r}^{2}-\lambda _{i}^{2} 
\end{eqnarray}

\begin{eqnarray}
  && \underset{h\to 0}{\mathop{\lim }}\,\frac{1}{h}\left( {{q}_{1}}+{{q}_{2}}-2 \right) \nonumber\\
 && =\underset{h\to 0}{\mathop{\lim }}\,\frac{\arctan \left( \sinh \left( h{{\lambda }_{1}} \right) \right)+\arctan \left( \sinh \left( h{{\lambda }_{2}} \right) \right)}{h} \nonumber\\ 
 && =\underset{h\to 0}{\mathop{\lim }}\,\frac{{{\lambda }_{1}}\cosh \left( h{{\lambda }_{1}} \right)}{1+{{\sinh }^{2}}\left( h{{\lambda }_{1}} \right)}+\frac{{{\lambda }_{2}}\cosh \left( h{{\lambda }_{2}} \right)}{1+{{\sinh }^{2}}\left( h{{\lambda }_{2}} \right)} \nonumber\\ 
 && ={{\lambda }_{1}}+{{\lambda }_{2}}
\end{eqnarray}

\begin{eqnarray}
  && \underset{h\to 0}{\mathop{\lim }}\,\frac{1}{{{h}^{2}}}\left( {{q}_{1}}+{{q}_{2}}-{{q}_{1}}{{q}_{2}}-1 \right) \nonumber\\ 
 && =\underset{h\to 0}{\mathop{\lim }}\,-\frac{\arctan \left( \sinh \left( h{{\lambda }_{1}} \right) \right)\arctan \left( \sinh \left( h{{\lambda }_{2}} \right) \right)}{{{h}^{2}}} \nonumber\\
 && =\underset{h\to 0}{\mathop{\lim }}\,-\frac{{{\lambda }_{1}}\operatorname{sech}\left( h{{\lambda }_{1}} \right)\arctan \left( \sinh \left( h{{\lambda }_{2}} \right) \right)+{{\lambda }_{2}}\arctan \left( \sinh \left( h{{\lambda }_{1}} \right) \right)\operatorname{sech}\left( h{{\lambda }_{2}} \right)}{2h} \nonumber\\
 && =\underset{h\to 0}{\mathop{\lim }}\,-\frac{\begin{split}
  & -\lambda _{1}^{2}\tanh \left( h{{\lambda }_{1}} \right)\operatorname{sech}\left( h{{\lambda }_{1}} \right)\arctan \left( \sinh \left( h{{\lambda }_{2}} \right) \right)+2{{\lambda }_{1}}{{\lambda }_{2}}\operatorname{sech}\left( h{{\lambda }_{1}} \right)\operatorname{sech}\left( h{{\lambda }_{2}} \right) \\ 
 & -\lambda _{2}^{2}\arctan \left( \sinh \left( h{{\lambda }_{1}} \right) \right)\tanh \left( h{{\lambda }_{2}} \right)\operatorname{sech}\left( h{{\lambda }_{2}} \right) \\ 
\end{split}}{2} \nonumber\\
 && =-\lambda_{1}\lambda_{2} 
\end{eqnarray}


\end{document}